\documentclass{aa}  

\usepackage{graphicx,subfigure}
\usepackage{txfonts}
\usepackage{soul} 
\usepackage{color}

\definecolor{purple}{rgb}{0.2,0.0,0.4}
\newcommand{\FG}[1]{{\color{black}{#1}}}

 \usepackage{mathrsfs}
 
 \newcommand{\G}{\mathcal{G}}
\newcommand{\Msun}{M_\odot}
\newcommand{\Rsun}{R_\odot}
\newcommand{\Ms}{M_{\star}}
\newcommand{\Rs}{R_{\star}}
\newcommand{\Ts}{T_\star}
\newcommand{\Mp}{M_{\rm p}}

\newcommand{\dd}{\mathrm{d}}
\newcommand{\Os}{\Omega_{\star}}

\newcommand{\Qs}{\overline{Q_s'}}

\begin{document}

   \title{Tidal dissipation in rotating low-mass stars \\ and implications for the orbital evolution of close-in planets}

   \subtitle{I. From the PMS to the RGB at solar metallicity}

\author{F. Gallet\inst{1} \and E. Bolmont\inst{2,3}  \and S. Mathis\inst{2} \and   C. Charbonnel\inst{1,4} \and L. Amard\inst{1,5} }

\offprints{F. Gallet,\\ email: florian.gallet@unige.ch}

\institute{ 
$^1$ Department of Astronomy, University of Geneva, Chemin des Maillettes 51, 1290 Versoix, Switzerland \\
$^2$ Laboratoire AIM Paris-Saclay, CEA/DRF - CNRS - Univ. Paris Diderot - IRFU/SAp, Centre de Saclay, F-91191 Gif-sur-Yvette Cedex, France \\
$^3$ NaXys, Department of Mathematics, University of Namur, 8 Rempart de la Vierge, 5000 Namur, Belgium \\
$^4$ IRAP, UMR 5277, CNRS and Universit\'e de Toulouse, 14, av. E. Belin, F-31400 Toulouse, France \\
$^5$ LUPM UMR 5299 CNRS/UM, Universit\'e de Montpellier, CC 72, F-34095 Montpellier Cedex 05, France
  }

   \date{Received --; accepted --}

  \abstract
 {Star-planet interactions must be taken into account in stellar models to understand the dynamical evolution of close-in planets. The dependence of the tidal interactions on the structural and rotational evolution of the star is of peculiar importance and should be correctly treated.}
  {We {quantify} how tidal dissipation in the convective envelope of rotating low-mass stars evolves from the pre-main sequence up to the red-giant branch depending on the initial stellar mass. We investigate the consequences of this evolution on planetary orbital evolution.} 
  {We couple the {tidal dissipation formalism} described in Mathis (2015) to the stellar evolution code STAREVOL {and apply it to} rotating stars {with masses between} 0.3 and 1.4 $M_{\odot}$. {As a first step, this formalism assumes a simplified bi-layer stellar structure with corresponding averaged densities for the radiative core and the convective envelope. We use a frequency-averaged treatment of the dissipation of tidal inertial waves in the convection zone (we neglect the dissipation of tidal gravity waves in the radiation zone).} In addition, we generalize the work of Bolmont \& Mathis (2016) by following the orbital evolution of close-in planets using {the} new tidal dissipation predictions for advanced phases of stellar evolution.} 
   {On the pre-main sequence the evolution of tidal dissipation is controlled by the evolution of the internal structure of the contracting star. On the main-sequence it is strongly {driven} by the variation of surface rotation {that is} impacted by {magnetized stellar winds braking}. The main effect of {taking into account the rotational evolution of the stars} is to lower the tidal dissipation strength by about four orders of magnitude on the main-sequence, compared to a normalized dissipation rate that only takes into account structural changes.}
{The evolution of the dissipation strongly depends on the evolution of the internal structure and rotation of the star. From the pre-main sequence up to the tip of the red-giant branch, it varies by several orders of magnitude, with strong consequences for the orbital evolution of close-in massive planets. These effects are the strongest during the pre-main sequence, implying that the planets are mainly sensitive to the star's early history.}

   \maketitle
%

\section{Introduction}

Thanks to space observatories and to the increase of the precision of modern techniques (e.g. radial velocity and transit methods), we have now access to a huge number of exoplanets that belong to a wide variety of star-planet systems configurations in which host stars are ranging from M red dwarf to intermediate-mass A-type stars \citep{Fabrycky14}. Among these discovered exoplanets, a fairly large number of them are found close to their host star as it is the case for the well known hot Jupiter's class exoplanet \citep{MQ95,Charbonneau2000}.

The presence of planets is usually not taken into account in the numerical codes dealing with the evolution of stellar rotation (such as angular momentum evolution codes, e.g. \citealt{RM12,GB13,GB15,Johnstone15,Lanzafame15}, or stellar evolution codes including angular momentum transport, e.g. \citealt{ES76,ES81,Pinsonneault90,Brott11,Amard15,Choi16}). But with the increasing number of detected and confirmed exoplanets, {especially the fact that most of them are found close to their host star,} star-planet interactions should not be neglected anymore (as shown by the studies of \citealt{Strugarek14} and \citealt{Strugarek16} for the magnetic interactions, and of \citealt{Bolmont16} for tidal interactions). \FG{Indeed, in those close-in configurations, the dissipation of tidal waves inside the turbulent convective envelope of low-mass stars is though to strongly affect the orbit of the surrounding planet \citep{Jackson2008,Husnoo12,Lai12,Guillot14}, the spin-orbit inclination \citep{Barker09,Winn10,Albrecht12}, as well as, in the case of a massive planet, the rotational evolution of the star \citep{Ogilvie07,Bolmont11,Bolmont12,Albrecht12,Ogilvie14,AD14,Mathis15,Bolmont16} and possibly its internal structure \citep{deBoer}.}

In stars, there are two components of tides: the equilibrium and the dynamical tides. 
The equilibrium tide {corresponds to a large scale hydrostatic adjustment of a body and the resulting flow due to the gravitational field of a given companion  {\citep{Zahn66,Remus2012}}. It is usually employed in the framework of the constant time lag model \citep[see][]{Mignard1979,Hut1981,Eggleton98,Bolmont11,Bolmont12}, which allows a fast computation of the orbital evolution of the planet and works for all eccentricities \citep{Hut1981,Leconte2010}}. In this model, the dissipation of the kinetic energy of the {equilibrium} tide inside the star is often taken to be constant throughout the system evolution and calibrated on observations \citep{Hansen10,Hansen12}. {While considering such a constant equilibrium tide dissipation is a sensible assumption, several studies showed that this quantity might vary during the different phases of stellar evolution. For example, \cite{ZahnBouchet1989} showed that the dissipation of the equilibrium tide by the turbulent friction in the convective envelope of late-type stars is strongest during their PMS. {Using this theoretical framework, \citet[][see also \citet{VP1995}]{Villaver09} recalled that the variation of the semi-major axis of a planet induced by such friction can be expressed as a function of the ratio of the mass of the convective envelope to the total mass of the star, the ratio between the radius of the star and the orbital semi-major axis (to the power 8), and finally of a power of the ratio between the tidal period and the convective turnover timescale. This allows one to model the loss of efficiency of tidal friction for rapid tides \citep[e.g.][]{Zahn66,GK1977}.}
Because of the variations of these quantities during post-MS phases \citep[e.g.][]{CC15}, this could lead to a more efficient dissipation than during the MS. Finally, \citet{Mathis16} demonstrated that the action of rotation on convection deeply modifies the turbulent friction it applies on the equilibrium tide. In the regime of fast rotation, which corresponds to the end of PMS and early MS phase, the friction is several orders of magnitude lower than in a model ignoring rotation. This may lead to a loss of efficiency of the dissipation of the equilibrium tide. This shows how one should be {careful} when assuming a calibrated constant dissipation of the equilibrium tide during the evolution of stars.}

{{On the other hand,} the dynamical tide corresponds to the excitation of tidal waves inside the star \citep{Zahn75,Ogilvie07}.}
In the dynamical tide formalism, the tidal dissipation in the  convective envelope of low-mass stars is due to the action of the convective turbulent friction applied on tidal inertial waves (mechanical waves that are generated inside rotating fluid bodies) driven by the Coriolis acceleration \citep{Ogilvie07,Mathis16}.
In the radiative {layers,} the dissipation is due to thermal diffusion and breaking mechanisms acting on gravito-inertial waves \citep[e.g.][]{Zahn75,Terquem98,Barker10}.

The properties of a star, {its} internal structure ({relative masses and radii of the} radiative core and convective envelope), and {its} rotation rate {strongly evolve along the stellar life}. The temporal evolution of the radius and mass of the radiative core and of the surface rotation rate has strong consequences on the evolution of the amplitude of the tidal dissipation in stars {along their evolution} \citep{Zahn66,Zahn75,Zahn77,Ogilvie07,Siessetal2013,Mathis15,Mathis16}. 
{This could explain} observations of star-planet and binary-star systems that show a range of tidal dissipation varying over several orders of magnitude, {as reported by \citet{Ogilvie14}.} Moreover, the tidal dissipation strongly impacts the dynamical evolution of planetary systems along the evolution of their host stars \citep[e.g.][]{Bolmont12,Bolmont16}. We thus need to take into account its potential variations as a function of stellar age using the best available ab-initio modeling.

{The work of \citet{Mathis15} constituted the first step towards a complete description of tidal dissipation along stellar evolution. Using a simplified two-layer model as in \citet{Ogilvie13}, \citet{Mathis15} followed the dissipation of the dynamical tide inside the convective envelope along the standard stellar evolution tracks of \citet{Siess2000}.}
\citet{Bolmont16} {then} included in their orbital evolution code  \citep{Bolmont11, Bolmont12} the prescription  {for the dynamical tide} of \citet{Mathis15} coupled to a simplified description of the evolution of the stellar surface rotation rate. This work led to the complete re-evaluation of the effects of star-planet tidal interactions on the orbital evolution of massive close-in planets. {In particular,} \citet{Bolmont16} reported outward and inward migrations of close-in hot Jupiters orbiting solar-type stars while no (or a small) evolution was initially found when including only the equilibrium tide component. While these {pioneer} developments represent an important step forward in understanding the complexity of the tidal interactions  between stars and planets, we now need to properly account for the evolution of the rotation of the star and its inter-connection with {secular} structure {variations}. 
{This is particularly important for characterizing the orbital evolution of} short period systems.
Actually, by using a constant (in time) quality factor \citep{Goldreich66} or time lag formalism, it is not possible to explain the hot Jupiter desert that we observe around rapidly rotating stars \citep{Lanza14,Teitler14,McQuillan13,Mazeh16}. 
Providing a simplified but realistic evolution of the tidal dissipation is crucial to predict the position at which planets are at any time and to rapidly explore the effects of {initial conditions} on their orbital evolution {and on possible planet engulfment that is expected to strongly affect the surface rotation of hot stars \citep{SiessLivio99,Privitera2016b,Privitera2016}}. {Such tools will be essential for} the preparation {and the exploitation of future observations with} CHEOPS \citep{CHEOPS}, TESS \citep{TESS}, and SPIRou \citep{Spirou}. 

{In this context, the originality of the present work is that we
{introduce for the first time in a stellar evolution code (STAREVOL) the prescription of \citet{Mathis15} for the dissipation of the dynamical tide inside stellar convective envelopes. This allows us to follow this quantity self-consistently together with the secular and rotational history of stars, from the pre-main sequence up to the red giant branch. We also take into account the equilibrium tide using the constant time lag model.}
This paper is organized as follows. In Sect. \ref{modeldes}, we recall the formalism and assumptions used to analytically express the frequency-averaged tidal dissipation \citep[see][]{Ogilvie13} and the micro physics and assumptions used in STAREVOL. In Sect. \ref{results} we describe the evolution of the dissipation as a function of mass, evolutionary phase, and rotation for stars ranging between 0.3 and 1.4 $M_{\odot}$. 
In Sect. \ref{orbit_evol}, we show the influence of the evolving structure and tidal dissipation on the orbital evolution of close-in planets around $1~M_\odot$ and $1.2~M_\odot$ stars. 
We conclude and discuss the perspectives of this work in Sect. \ref{conc}.

\section{Model description}
\label{modeldes}

\subsection{Tidal dissipation formalism}
\label{formalism}

\begin{figure}
\begin{center}
\includegraphics[angle=-90,width=0.5\textwidth]{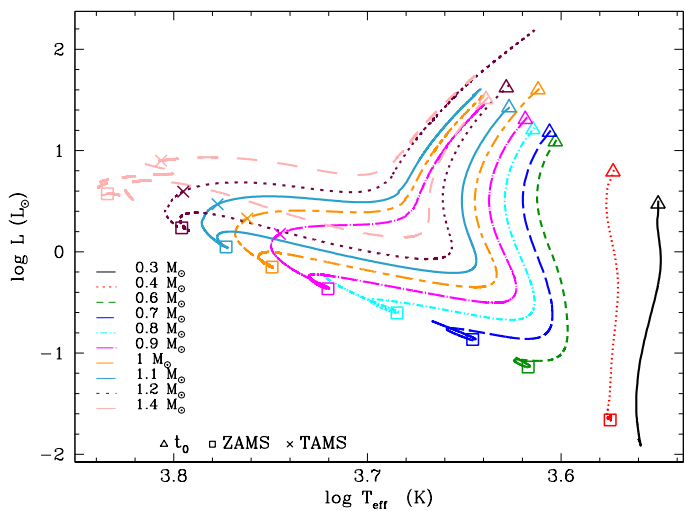}
\end{center}
\caption{Stellar evolution tracks in the Hertzsprung-Russell diagram for the rotating models of 0.3, 0.4, 0.6, 0.7, 0.8, 0.9, 1.0, 1.1, 1.2, and 1.4~$M_{\odot}$ at solar metallicity. {We show the evolution up to the RGB for the more massive stars or up to the evolution stage the models reach at 20~Gyr for the less massive stars.}
The symbols represent the first step in each evolution sequence (triangle), the ZAMS (square), and the TAMS (cross).}
\label{hrd}
\end{figure}
In this section we describe {the method we use} to couple, for the first time, the structural and rotational evolution of low-mass stars with tidal dissipation in their convective envelope. 

\subsubsection{Generalities}

\begin{figure*}[!ht]
\begin{center}
\includegraphics[angle=-90,width=0.5\textwidth]{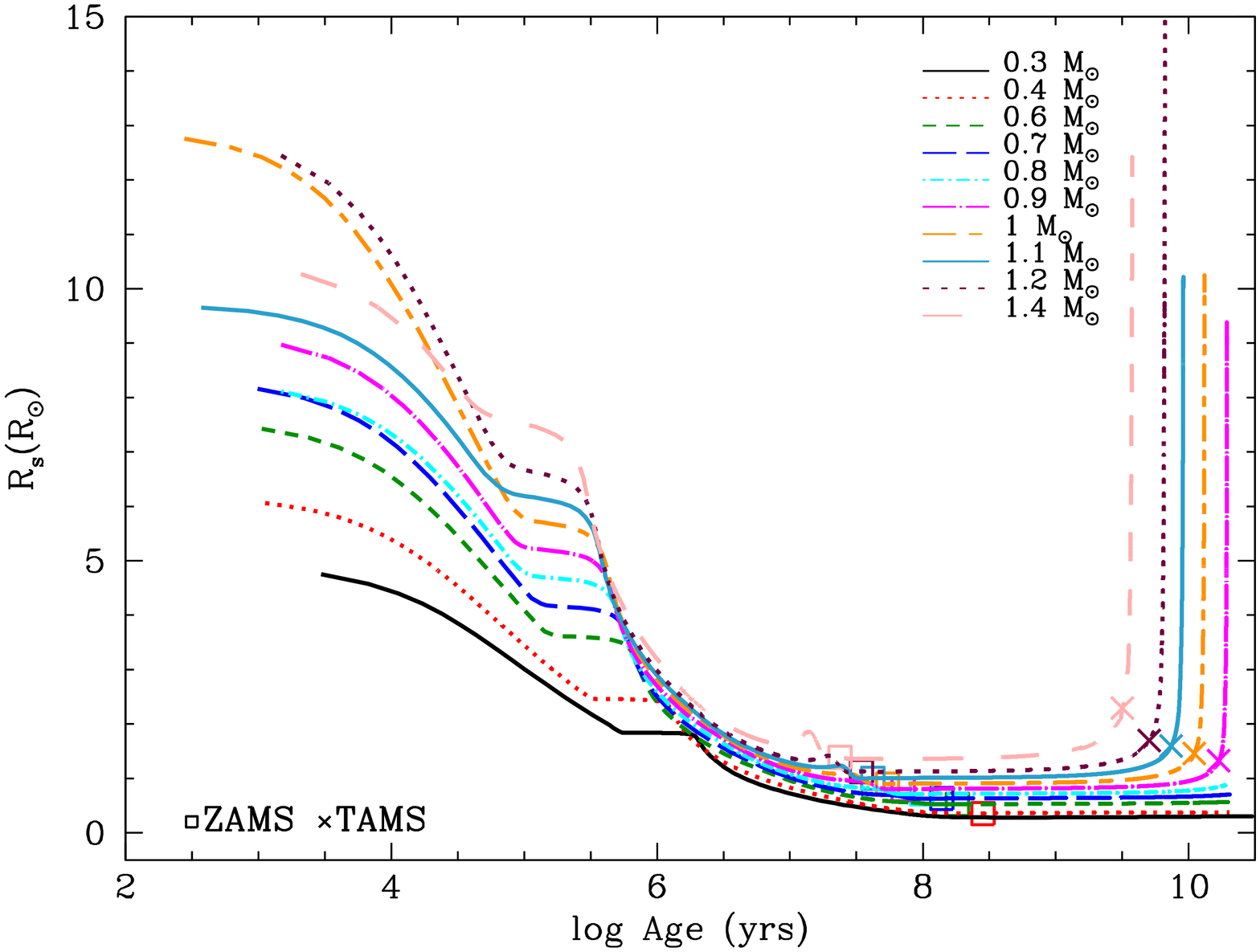} \hspace{-0.5cm} \includegraphics[angle=-90,width=0.5\textwidth]{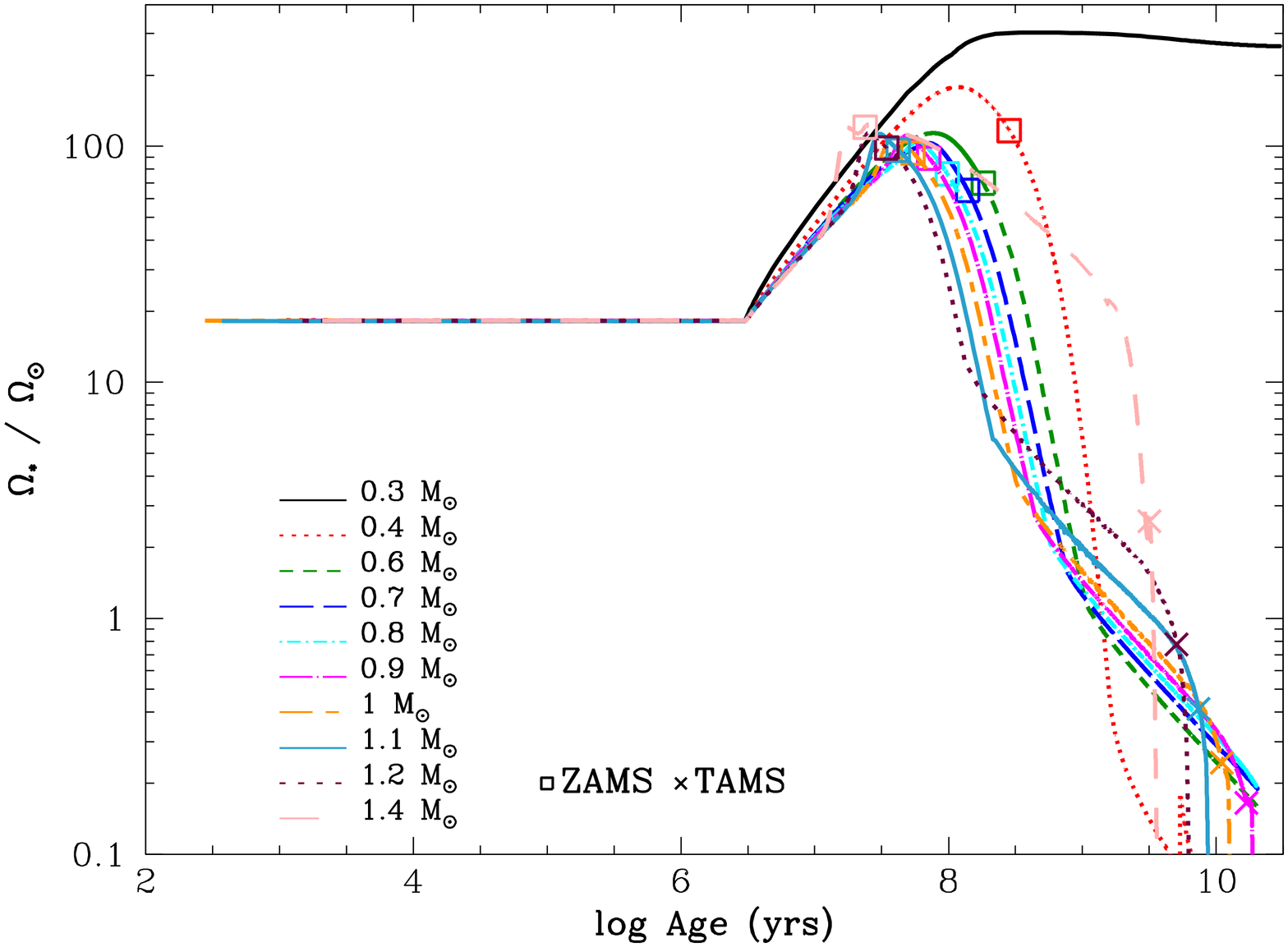}  \\
\includegraphics[angle=-90,width=0.5\textwidth]{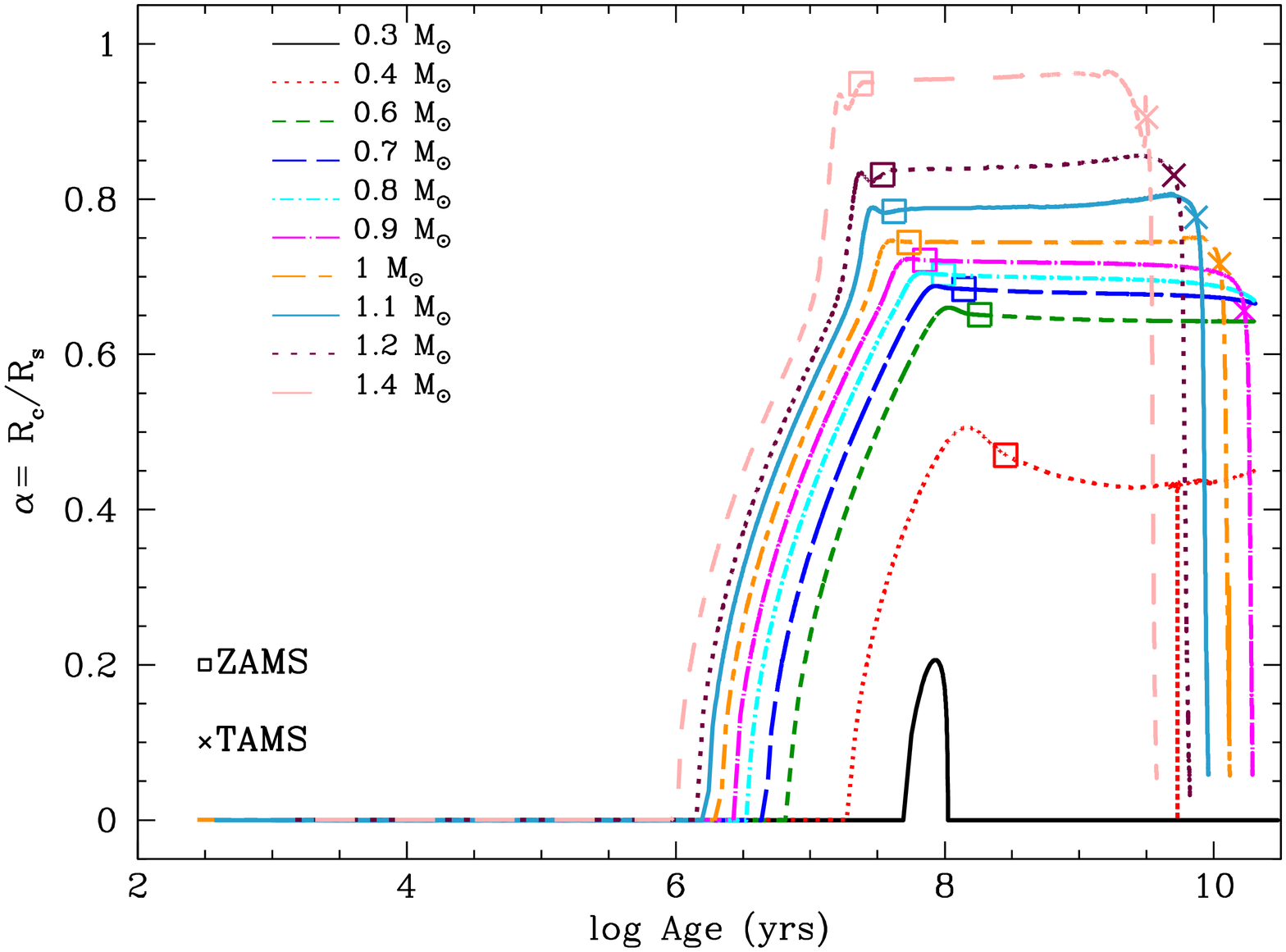} \hspace{-0.5cm} \includegraphics[angle=-90,width=0.5\textwidth]{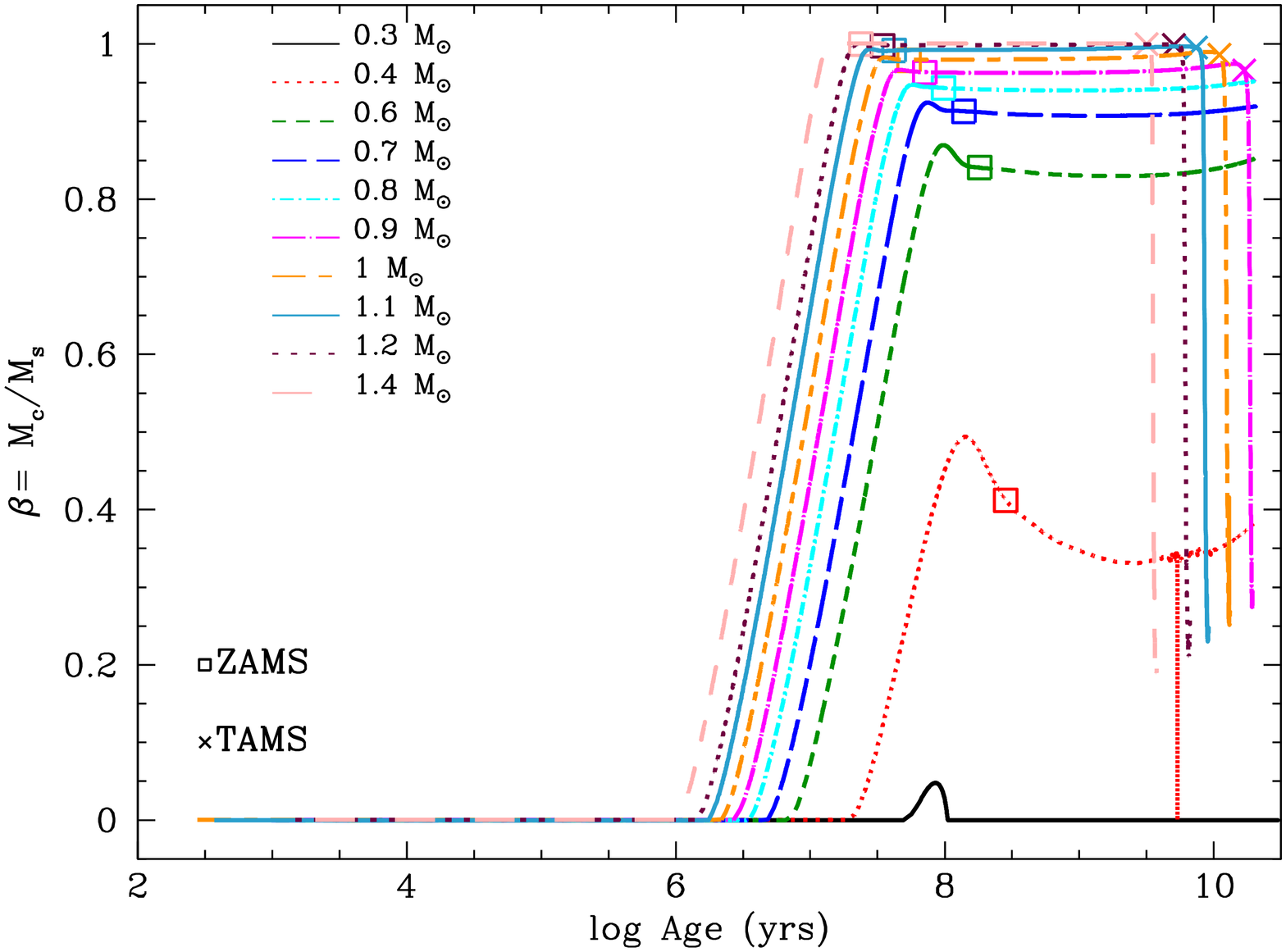}  
\end{center}
\caption{{Top-left:} Evolution of the stellar radius $R_{\star}$ of stars from 0.3 to 1.4 $M_{\odot}$ as a function of time. {Top-right:} Surface angular velocity (scaled to the present Sun angular velocity $\Omega_{\odot}=2.87 \times 10^{-6} $s$^{-1}$) evolution for the different stellar masses. {Bottom-left:} Evolution of the radius aspect ratio $\alpha=R_c/R_{\star}$ of stars from 0.3 to 1.4 $M_{\odot}$ as a function of time. {Bottom-right:} Evolution of the mass aspect ratio $\beta=M_c/M_{\star}$ of stars from 0.3 and 1.4 $M_{\odot}$ as a function of time. The symbols represent: the first step in each model (triangle), the ZAMS (square), and the TAMS (cross). Table \ref{table} summarize at which evolutionary step the models of this work stop.}
\label{stellarstructure}
\end{figure*}
Let us consider two bodies: a deformable star and a point-mass planet. The planet exerts a differential force on the star that causes {its deformation which leads to the generation of} tidal flows. These tidal flows are submitted to friction in its interior (as it is the case for the synchronization of massive binaries by the damping of gravity waves near the stellar surface; \citealt{Goldreich89}). Part of the kinetic energy associated to the flows is converted and lost in thermal energy inside the star and part of it is transferred to the planet's orbit via angular momentum exchanges. These processes are governed by the tidal dissipation which depends on the dissipative processes and the stellar internal structure.
The complex \FG{tidal} interactions between star and planets is decomposed into the equilibrium tide and the dynamical tide \citep[e.g.][]{Zahn66,Zahn75,Zahn77,MathisRemus13,Ogilvie14} {that are both considered in the present study}.

The equilibrium non-wave-like tide corresponds to the internal large-scale flows produced by the hydrostatic adjustment of the stellar structure due to the presence of a companion \citep{Zahn66,Remus2012,Ogilvie13}. This tide is efficiently dissipated in the convective envelope of rotating low-mass stars by turbulent friction due to convection motion \citep{Zahn66,Zahn89,Ogilvie12,Mathis16}. In most of the studies about tidal evolution of planetary systems, only the equilibrium tide is taken into account \citep[e.g.][]{Mignard1979,Hut1981,Leconte2010,Bolmont11}. In this work we {model this component using} the framework of the constant time lag model \citep[see][]{Mignard1979,Hut1981,Eggleton98,Bolmont11,Bolmont12}.

{On the other hand,} the dynamical tide comes from inertial waves propagating through the convective envelope that are driven by the Coriolis acceleration and excited when the tidal excitation frequency $|\omega|$ is smaller than $2\Omega_{\star}$, where $\Omega_{\star}$ is the stellar spin. The tidal frequency is defined in \citet{Ogilvie14} as the linear combination of the orbital and spin frequencies with small integer coefficients. If the star-planet system is coplanar and
the planet is on a circular orbit the tidal frequency can be expressed as $\omega \equiv 2(n-\Omega_{\star})$, where $n$ is the orbital frequency \citep{Ogilvie04}. For tidal frequencies $|\omega| > 2\Omega_{\star}$ the tidal dissipation is almost independent of $\Omega_{\star}$ at any given tidal frequency because in that regime the effect of the Coriolis force is weak \citep{Ogilvie07}. {In the radiative core the dynamical tide is {driven by} internal gravity waves \citep[see][]{Zahn75,Zahn77,Goldreich892,Goldreich89,Terquem98} which can be affected by the Coriolis acceleration \citep{Ogilvie07}.

{Both the equilibrium and the dynamical tidal effects} must {in principle} be accounted for} to properly model tidal evolution \citep{Bolmont16}. However, while the equilibrium tide weakly depends on the excitation frequency variation \citep{Remus2012}, the dynamical tide strongly depends on it, and also on the evolutionary stage, mass, and rotation rate of the star \citep[see the discussion in][]{Ogilvie04,Ogilvie07,Barker10,ADMLP14,Witte02}. 

The formalism associated to the dynamical tide \citep[see][]{Ogilvie13,Ogilvie14,Mathis15} is currently too {complex} to be implemented in secular orbital evolution codes \citep[see][]{Bolmont16} and to perform wide explorations {of the parameter space (planet and stellar masses, initial rotation and orbital configurations, along the whole stellar evolution)}. Indeed, the dynamical tide dissipation spectrum harbors complex behaviors, as {it} evolves as a function of the stellar properties and age, that are computer consuming. 
{A first step {to make significant progress} is to follow the evolution of the tidal dissipation using rotating stellar models, considering {first} the dissipation inside the convective envelope of the stars.} 

\subsubsection{Frequency-averaged tidal dissipation}

In the formalism of \citet{Ogilvie13} and \citet{Mathis15}, the stellar convective envelope is assumed to be in solid-body rotation with angular velocity $\Omega_{\star}$. {Moderate rotation is assumed, i.e.,} the squared ratio of $\Omega_{\star}$ to the critical angular velocity $\Omega_{\rm{c}}$ is such that $\left(\Omega_{\star}/\Omega_{\rm{c}} \right)^2 =  \left(\Omega_{\star}/ \sqrt[]{\mathcal{G} M_{\star}/R_{\star}^3} \right)^2 \equiv \epsilon^2 \ll 1$ \FG{(so as to neglect the centrifugal effect)}, where $\mathcal{G}$ is the gravitational constant, and $M_{\star}$ and $R_{\star}$ are the stellar mass and equatorial radius, respectively. 
In this article we use the two-layers model introduced in \citet{Ogilvie13} and \citet{Mathis15} to evaluate the frequency-averaged tidal dissipation in the {stellar} convective envelope, {and we focus on solar-metallicity} stars {with initial masses} between 0.3 and 1.4 $M_{\odot}$. In this mass range, the convective envelope surrounds the radiative core of radius $R_{\rm{c}}$ and mass $M_{\rm{c}}$. Both core and envelope are assumed to be homogeneous with respective average densities $\rho_{\rm{c}}$ and $\rho_{\rm{e}}$. {This constitutes a {necessary first} step that allows us to derive an analytical expression for the frequency-averaged dissipation and to explore a broad space of parameters. In {the} near future, we shall evaluate the impact of the radial variations of the density, which varies in stellar convection zones over several orders of magnitude during the evolution of stars. This may lead to {weaker} dissipation rates.}

In the case of a coplanar star-planet system in which the orbit of the planet is circular, the frequency-averaged tidal dissipation \citep{Ogilvie13,Mathis15} is given by:
\begin{flalign}
\label{dissipequa}
<\mathcal{D}>_{\omega} =& \int^{+\infty}_{-\infty} \rm{Im}\left[k_2^2(\omega)\right] \frac{\dd\omega}{\omega} = \frac{100\pi}{63}\epsilon^2 \left( \frac{\alpha^5}{1-\alpha^5} \right)  \left( 1-\gamma \right)^2 \\ 
\times&  \left( 1-\alpha \right)^4  \left( 1 +2\alpha+3\alpha^3 + \frac{3}{2}\alpha^3 \right)^2 \left[ 1+  \left( \frac{1-\gamma}{\gamma} \right) \alpha^3  \right] \nonumber \\ 
\times& \left[   1 + \frac{3}{2}\gamma + \frac{5}{2\gamma}  \left( 1 + \frac{1}{2}\gamma - \frac{3}{2}\gamma^2 \right)  \alpha^3 - \frac{9}{4} \left( 1-\gamma \right)\alpha^5   \right]^{-2} \nonumber,
\end{flalign}
with 
\begin{flalign}
\alpha = \frac{R_{\rm{c}}}{R_{\star}}, {\rm~} \beta=\frac{M_{\rm{c}}}{M_{\star}}, {\rm~} \gamma = \frac{\rho_{\rm{e}}}{\rho_{\rm{c}}}=\frac{\alpha^3(1-\beta)}{\beta(1-\alpha^3)} < 1.
\end{flalign}
$k_2^2$ is the Love number of degree 2 corresponding to the quadrupolar mode ($k_l^m$, with $l$ = 2 and $m$ = 2 the components of the time-dependent tidal potential proportional to the spherical harmonic $Y_l^m$) that gives the ratio between the perturbation of the gravitational potential induced by the presence of the planetary companion and the tidal potential evaluated at the stellar surface. Its imaginary component $\rm{Im}\left[k_2^2(\omega)\right]$ is a direct estimation of the tidal dissipation.
The interest of this formalism is that it is possible to decompose Eq. \ref{dissipequa} into two parts: the factor $\epsilon^2$ on the one hand, and the part of Eq.~\ref{dissipequa} that is a unique function of $\alpha$ and $\beta$ on the other hand. The first part takes into account the rotation rate of the star (via $\epsilon$) and the second part only takes into account the dependence on the internal stellar structure (via the structural parameters $\alpha$ and $\beta$).
As in \cite{Mathis15}, we can therefore express the frequency-averaged dissipation at a fixed rotation: 
\begin{eqnarray}
< \mathcal{D}>_{\omega}^{\Omega} =  \epsilon^{-2} < \mathcal{D}>_{\omega}  =    \epsilon^{-2} < \rm{Im}\left[k^2_2(\omega)\right] >_{\omega},
\end{eqnarray}
that only depends on $\alpha$ and $\beta$. 
We can also define a second frequency-averaged dissipation using the critical angular velocity of the Sun $\Omega^{\odot}_{\rm{c}}$ instead of that of the star: 
\begin{eqnarray}
\label{norm_sun}
<\hat{\mathcal{D}}>_{\omega}^{\Omega}=\hat{\epsilon}^{-2} <\mathcal{D}>_{\omega} = \left( \frac{M_{\star}}{M_{\odot}} \right)^{-1}  \left( \frac{R_{\star}}{R_{\odot}} \right)^{3} < \mathcal{D}>_{\omega}^{\Omega},
\end{eqnarray}
where $\hat{\epsilon}^{2} \equiv \left(  \Omega_{\star} / \sqrt[]{\mathcal{G}M_{\odot}/R_{\odot}^3}  \right)^2 = \left( \Omega_{\star}/\Omega^{\odot}_{\rm{c}} \right)^2$ with $M_{\odot}$ and $R_{\odot}$ the mass and radius of the Sun that allow us to express the variation of the radius of the star along time. {The frequency-averaged dissipation provides us a reasonable order of magnitude of the friction applied on tidal inertial waves in a rotating convective envelope as a function of its structural properties (radius and mass aspect ratios) and rotation rate. However, \citet{Ogilvie07} showed how the dissipation of these waves can vary over several orders of magnitude {when inertial waves are excited}. This may lead, for a given frequency, to strong differences with the frequency-averaged value. Taking into account such a complex frequency-dependence would require to couple coherently high-resolution hydrodynamical numerical simulations of tidal inertial waves with secular stellar evolution and orbital codes and to heavy computation procedures \citep[e.g.][]{Witte02}. Using, as a first step, frequency-averaged dissipation thus constitutes an intermediate and necessary step that allows us to explore a broad {parameter space} for planetary systems and their host stars.}

\subsubsection{Modified equivalent tidal quality factor}

In a large number of studies about the tidal evolution of planetary systems, a quantity called the equivalent tidal quality factor is used \citep[e.g.][]{Goldreich66}. This quantity comes from the modeling of the tidal response with an idealized system made of a harmonic oscillator (the forcing frequency corresponds to the excitation frequency imposed by the perturbing body and the Coriolis acceleration is the restoring force) and a damper  \citep[corresponding in this framework to a turbulent viscosity, see][]{Greenberg2009}. {Following \citet{Ogilvie07}}, the equivalent {modified} tidal quality factor $\overline{Q'}$ {is introduced} and expressed in terms of the tidal dissipation $<\mathcal{D}>_{\omega}$ as:
\begin{eqnarray}
\label{qnonormequa}
\overline{Q'} = \frac{3}{2 <\mathcal{D}>_{\omega}}=\frac{3}{2}\frac{\overline{Q}}{k_2}.
\end{eqnarray}
{In this equation, we also recall the usual expression as a function of the equivalent tidal quality factor $\overline{Q}$ and the second-order Love number $k_2$. For an homogeneous fluid body $k_2 = 3/2$ and $\overline{Q'} = \overline{Q}$. Using $\overline{Q'}$ allows us to avoid to compute explicitely $k_2$. Indeed, we recall that the real physical quantity is the dissipation while its expression as a function of $k_2$ and ${\overline Q}$ comes from the simplified constant tidal quality factor model \citep[e.g.][]{MacDonald1964}.} {By definition, the lower the equivalent tidal quality factor, the more thermal energy is liberated into the star by the tidal dissipation process, and the {stronger} the impact on the planet's orbit.}

The modified equivalent tidal quality factor is usually considered as a free parameter to fit to a given star-planet system \citep{Jackson2008,FerrazMello2015}. Moreover, this quantity is often assumed to be constant throughout the entire stellar evolution \citep[e.g.,][]{MardlingLin2002,Jackson2008,FerrazMello2015}, {which leads to non realistic orbital evolution that cannot reproduce the observed dearth of hot-Jupiter around rapidly rotating stars \citep{Lanza14,Teitler14,McQuillan13,Mazeh16}}. 

Just as we could estimate the impact of the stellar structure on tidal dissipation (i.e., the tidal dissipation at a fixed rotation rate), we can define the equivalent structural quality factor as follows:
\begin{eqnarray}
\label{qnormequa}
\overline{Q'}_s = \hat{\epsilon}^2\overline{Q'}=   \frac{3}{2} \frac{\hat{\epsilon}^2}{<\mathcal{D}>_{\omega}} = \frac{3}{2} \frac{1}{<\mathcal{\hat{D}}>_{\omega}^{\Omega}}.
\end{eqnarray}

\subsubsection{Orbital evolution model}\label{orb_evol_model}

In order to compute the orbital evolution of close-in planets we use the model introduced in \citet{Bolmont16}. The evolution of the semi-major axis $a$ of a planet on a circular orbit is given by \citep{Hansen10,Leconte2010,Bolmont11,Bolmont12}: 
\begin{equation}\label{Hansena}
\frac{1}{a}\frac{\dd a}{\dd t} = - \frac{1}{\Ts}\Big[1-\frac{\Os}{n}\Big],
\end{equation}
where $n$ is the orbital frequency of the planet, and $\Ts$ is an evolution dissipation timescale given by:
\begin{equation}
\label{Ts}
\mathbf{\Ts = \frac{2}{9}\frac{\Ms}{\Mp(\Mp+\Ms)}\frac{a^8}{\Rs^{5}}\frac{\overline{Q'}_s}{\hat{\epsilon}^2}\frac{|n-\Os|}{\G}},
\end{equation}
which depends on the semi-major axis $a$ of the planet, the mass $\Ms$ and radius $\Rs$ of the star, the mass $\Mp$ of the planet, {the stellar equivalent structural quality factor $\overline{Q'}_s$}, and {$\G$ the gravitational constant}.  
Eq.~\ref{Ts} shows that: 1) the farther the planet, the higher the evolution timescale, 2) the smaller the radius of the star, the higher the evolution timescale, 3) {the bigger the quality factor $\overline{Q'}_s$, the higher the evolution timescale}.
{We would like to point out here a typo in \citet{Bolmont16}, where a $k_2$ factor was forgotten in Eqs. (4) and (10)\footnote{{Eq.~4 of \citet{Bolmont16} is $k_2/\overline{Q} = \sin\left[2\delta\right]$, but it should be $k_2/\overline{Q} = k_2\sin\left[2\delta\right]$ \citep{Remus2012}. This leads to $\Delta\tau_\star = \frac{3}{4k_2\overline{Q'}|n-\Os|}$.}}. Despite the typo, the numerical results of \citet{Bolmont16} are however correct.}

{As in \citet{Bolmont12} and \citet{Bolmont16}, we consider the influence of tides and of the stellar wind on the rotation of the star.
The expression for the angular momentum loss rate is from {the modified \citet{Kawaler88}'s braking law proposed in \citet{Bouvier1997}}: 
\begin{align} \label{wind}
\begin{split}
  \frac{1}{J}\frac{\dd J}{\dd t} &= \frac{-1}{J}K\Omega_\star^\mu \omega_{\rm sat}^{3-\mu}\left(\frac{R_{\star}}{\Rsun}\right)^{1/2}\left(\frac{M_{\star}}{\Msun}\right)^{-1/2} \\
 & + \frac{1}{J}\frac{h}{2T_\star}\left[1-\frac{\Omega_{\star}}{n}\right],
\end{split}
\end{align}
where $J$ and $h$ are the stellar and orbital angular momentum, respectively. The parameters $K$, $\mu$ and $\omega_{\rm sat}$ are wind parameters of the model from \citet{Bouvier1997}. We refer the reader to \citet{Bolmont16} for the values of these parameters.} {We also recall the reader that the braking law used in this orbital evolution model is somewhat outmoded compared to the recent theoretical advances in this field \citep[see][]{Matt15,Reville15}. While including a more realistic braking law will not affect the general conclusion of this work, it could lead to small deviation in a given orbital evolution. Conversely, and since the star is considered as a solid body in this work, including the core-envelope decoupling in the model will lead to very distinct orbital evolution. In this framework, we are now investigating the effect of a more realistic rotational evolution on the orbital evolution of close-in planets that will be presented in a forthcoming paper.}


{When the dynamical tide is driving the evolution, the structural equivalent tidal quality factor ($\overline{Q'}_s$) is given by Eq. (\ref{qnormequa}). When the equilibrium tide is driving the evolution, we use the observational constraints of \citet{Hansen12}, which are given in terms of {a constant normalized tidal dissipation factor $\overline{\sigma_{\star}}$, depending on the stellar mass}. We refer to \citet{Bolmont16} for the correspondence between the tidal quality factor and the tidal dissipation factor $\overline{\sigma_\star}$.
{For instance, the normalized dissipation factor for a $1.0~\Msun$ star is taken to be $\overline{\sigma_\star} = 3\times 10^{-7}$ and for a $1.2~\Msun$ star it is $\overline{\sigma_\star} = 7.8\times 10^{-8}$.} 
We recall that assuming a constant dissipation of the equilibrium tide constitutes a simplified model that should be improved in a near future. Indeed, as explained in the introduction, it varies along the evolution of stars \citep[e.g.][]{ZahnBouchet1989,Villaver09,Mathis16}.}

~~\\
\noindent {While the work of \citet{Mathis15} provides a realistic evaluation of the evolution of the tidal dissipation for low-mass stars {from the pre-main sequence (PMS) to the subgiant (SG) phase}, it was done at fixed  {stellar rotation along the evolution}. 
{Here we} go one step forward and treat rotation evolution coherently {in STAREVOL. This allows us} {to follow the impact of rotation on the stellar structure and evolution tracks, and} to study {self-consistently} the dissipation of the tidal waves inside the convective envelope of rotating stars over a larger range of evolutionary phases.} 

\subsection{{Models of low-mass stars including rotation}}
\label{model}

This study is based on a grid of {stellar models of rotating stars} {we} computed with the code STAREVOL \citep[see e.g.][]{Amard15} for a range of initial masses between 0.3 and 1.4~M$_{\odot}$ at solar metallicity (Z = 0.0134; \citealt{AsplundGrevesse2009}). Figure \ref{hrd} shows the stellar evolution tracks of these {models} in the Hertzsprung-Russell {diagram}. Table \ref{table} summarize at which evolutionary step the models of this work stop.

The {references for the} basic input micro-physics (equation of state, nuclear reactions, and opacities) can be found in \cite{Amard15} and in \cite{Lagardeetal12}. The initial {helium} abundance and mixing length parameter are calibrated without {atomic}  diffusion to reproduce a non-rotating Sun with respect to the solar mixture of \cite{AsplundGrevesse2009} with a $10^{-5}$ precision for the luminosity and the radius at the age of the Sun. The corresponding mixing length parameter and initial helium abundance are $\alpha_{\rm{MLT}} = 1.6267$ and $Y = 0.2689$.

{The stellar evolution models are computed taking into account rotation. More specifically :}
\begin{enumerate}
\item[--] The evolution of angular momentum {in the stellar interior} is calculated from the first iteration step on the PMS phase {and up to the red-giant branch (RGB) following} the formalism developed by \citet{Zahn92}, \citet{MaeZah98} and \citet{MaZa04}. This formalism takes into account advection by meridional circulation and diffusion by shear turbulence \citep[see][]{Palacios03,Palacios06,Decressin09}. {The internal transport prescriptions used to describe turbulent diffusion coefficients are \citet{MPZ04} in the horizontal direction and \citet{TZ97} in the vertical one.}

\item[--] The convective region is assumed to be in solid-body rotation and is subject to magnetic braking from the PMS onward and up to the RGB following \citet{Matt15} prescription. The mass loss rate is estimated using the prescription of \citet{Cranmer11}.
 
\item[--] Star-disc interaction is taken into account during the early-PMS phase (i.e. during the 2 to 10 first Myr). Following \citet{GB15} surface rotation rate is assumed to be held constant during a characteristic timescale (the disk's lifetime). This phase is considered as an initial condition for angular momentum evolution and is fixed by the observations \citep[see][]{GB15}. 

\item[--] The initial stellar rotation is fixed using the calibration for fast rotators from \citet[][]{GB15}: an initial rotation period of 1.4 days (corresponding to $\Omega_{\star}=18~\Omega_{\odot}$) and a disk's lifetime of 3 Myr corresponding to the calibration of the solar-type stars. We applied this parameterization to the whole range of masses ($0.3$ to $1.4~M_\odot$) to analyze the impact of the stellar mass on the evolution of the dissipation. To reproduce the observed distribution of surface rotation period in star-forming regions and young-open clusters we should have calibrated the initial conditions {for each stellar mass} \citep{GB13,GB15,Amard15}. However, this calibration is out of the scope of this present study where we perform a wide exploration of the parameter space.   
\end{enumerate}

Figure~\ref{stellarstructure} shows the evolution {as a function of time} of the main stellar quantities {that enter in the expression of the equivalent tidal quality factor, namely,} the stellar radius ($R_{\star}$), the surface angular velocity ($\Omega_{\star}$), the radius {aspect} ratio $\alpha = R_c/R_{\star}$, and the mass {aspect} ratio $\beta=M_c/M_{\star}$. {This is shown for all the stellar masses considered in the computations.}

\FG{Note that our stellar evolution and orbital evolution models are not strictly coupled. Grid of structural quality factor $\Qs$ are initially computed using STAREVOL for stars with an initial rotation period of 1.4 days and latter provided to the orbital evolution code of \citet{Bolmont12} {with which we} compute the rotational evolution of the stars including the tidal torque and the torque produced by the stellar winds (see Eq. \ref{wind}).}
\begin{table}[ht]
\caption{\label{table} Phases and ages reached by our models at the end of each of the simulations.}
\center
   \begin{tabular}{c c c}
   \hline
   $M_{\star}$ & Phase & Age \\
   \hline
0.3 $M_{\odot}$ & PMS & 30.21 Gyr \\
0.4 $M_{\odot}$ & MS & 20.17 Gyr \\
0.6 $M_{\odot}$ & MS & 19.98 Gyr \\
0.7 $M_{\odot}$ & MS & 20.49 Gyr \\
0.8 $M_{\odot}$ & MS & 19.99 Gyr \\
0.9 $M_{\odot}$ & RGB & 19.29 Gyr \\
1.0 $M_{\odot}$ & RGB & 13.05 Gyr \\
1.1 $M_{\odot}$ & RGB & 9.12 Gyr \\
1.2 $M_{\odot}$ & RGB & 6.59 Gyr \\
1.4 $M_{\odot}$ & RGB & 3.73 Gyr \\
\hline
   \end{tabular}
\end{table}

\section{Tidal dissipation along the evolution of rotating stars}
\label{results}

{As described by Eq.~\ref{dissipequa}, the frequency-averaged tidal dissipation intrinsically follows the evolution of both {the} internal structure (through the $\alpha$ and $\beta$ parameters and $R_{\star}$) and {the} rotation rate (through the $\epsilon$ parameter) {of the star}. Here we investigate successively the effect of the evolution the stellar structure {(\S~\ref{structuraleffects})} and of the rotation rate {(\S~\ref{rotationaleffects})} on the evolution of the frequency-averaged tidal dissipation and equivalent modified quality factor.} 

\subsection{Structural effect: {evolution paths in the ($\alpha$, $\beta$) plane}}
\label{structuraleffects}

{To analyze how the dissipation evolves with the stellar structure, we first consider the frequency-averaged dissipation at fixed normalized angular velocity as in \citet{Mathis15}\footnote{\label{footnote1}{The evolution models of \citet{Siess2000} used by \citet{Mathis15} were computed without rotation. However in the case of low-mass stars, rotation has only a very modest effect on the stellar tracks (effective temperature and luminosity) and on the stellar structure (aspect ratios). We thus expect to find very similar behaviors compared to \citet{Mathis15} predictions.}}.}
Figure \ref{chemin} shows (color coded gradient) the intensity of the frequency-averaged dissipation at fixed normalized angular velocity ($<\mathcal{D}>_{\omega}^{\Omega}$) as a function of both mass and radius aspect ratios ($\alpha$ and $\beta$). The right-hand lower white region of Fig.~\ref{chemin} {is excluded as it is the} non physical ($\alpha$, $\beta$) area where the condition $\gamma < 1$ {(i.e. the core denser than the envelope)} is not fulfilled. 
$<\mathcal{D}>_{\omega}^{\Omega}$ exhibits a maximum in an island region around ($\alpha_{\rm{max}}=0.572, \beta_{\rm{max}}=0.503$) corresponding to an intensity of $1.091~10^{-2}$. Note that the regimes where stars are almost fully convective correspond to regular inertial waves for which dissipation is weak \citep[][{; black and dark blue colors}]{Wu05}, while when the radiative core is sufficiently extended sheared wave attractors with strong dissipation can form \citep[][{yellow to white colors}]{Ogilvie07}.

The interest is then to {overplot} in the ($\alpha$, $\beta$) parameter space  the evolution tracks of our stellar models (including rotation, but see footnote \ref{footnote1}) {and to describe their behavior at each evolution phase.} 

\begin{figure}[!ht]
\includegraphics[angle=-90,width=0.5\textwidth]{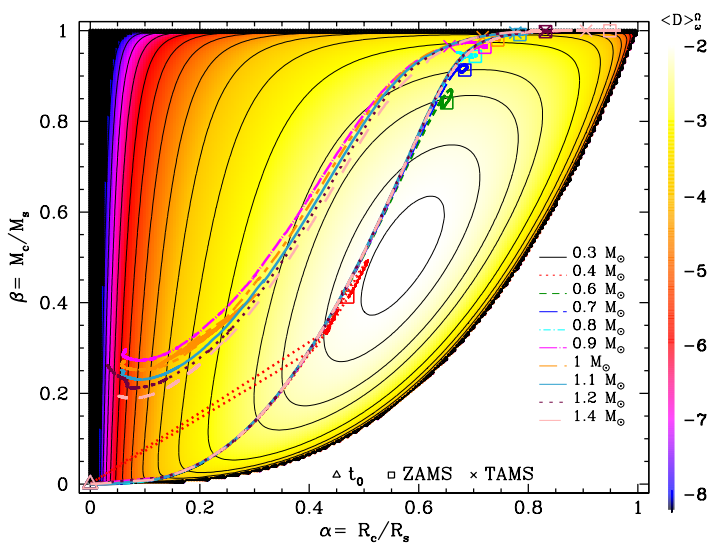}
\caption{Variation of normalized dissipation $<\mathcal{D}>_{\omega}^{\Omega}$ as a function of {the} radius and mass aspect ratios ($\alpha$ and $\beta$, respectively) in color scale. Levels are for $\log <\mathcal{D}>_{\omega}^{\Omega}=$ \{-2, -2.1, -2.3, -2.5, -3, -3.5, -4, -4.5, -5, -5.5, -6, -6.5, -7,
-7.5, and -8\}. The evolutionary {paths} {of the stellar models of the different masses (see labels)} {are overploted} in the ($\alpha, \beta$) plane. {Symbols are the same as in Fig.~\ref{hrd}.}}
\label{chemin}
\end{figure}

\subsubsection*{PMS-ZAMS}

{{Since all the models start their evolution with a fully convective interior, $\alpha$ and $\beta$ are initially equal to zero. Then both aspect ratios increase as the star contracts and the radiative core develops during the PMS up to the zero-age main sequence (ZAMS).}}
{{As a consequence,} the general behavior of the tidal dissipation {shown in Fig.~\ref{chemin}} can be easily understood. As both mass and radius aspect ratios of the models increase along the PMS}, {the stars successively pass through regions of increasing intensity until {they} brush against the islet of maximum intensity (surrounded by the level log $<\mathcal{D}>_{\omega}^{\Omega} = -2$, see Fig.~\ref{chemin}).} According to Fig. \ref{stellarstructure}, {the stars with masses higher than 0.3 $M_{\odot}$ reach this islet in a short timescale} between 3 Myr (1.4 $M_{\odot}$) and 100 Myr (0.4 $M_{\odot}$). {Then they move away from this maximum intensity region as both mass and radius aspect ratios continue to increase {while the stars approach} the ZAMS.} 
{Just before the ZAMS, several reactions have already been initiated such as the p-p chain and the CN reaction that produce enough energy to stop the stellar contraction. {The more massive stars then develop a convective core while their effective temperature and luminosity slightly decrease as they settle on the ZAMS.} This affects the whole radius of the star, {explaining the ``bump"} that is clearly visible in Fig. \ref{hrd} {for all the quantities, and inducing a sharp increase of the dissipation}. 
This increase is not visible in Fig. \ref{chemin} but clearly visible {in the left panels of Fig. \ref{dissip} for the highest mass star}. The case of the 0.3 $M_{\odot}$ star is quite interesting because it {hosts} a very small radiative core during a very brief moment ($\Delta t = $ 50 Myr) before reaching the ZAMS. This stellar mass represents the limit between fully convective and {partly radiative} stars. {{As pointed before, the dynamical tide induced dissipation is lower for fully convective stars} because the tidal waves that propagate through the convective envelope of the star require the presence of a radiative core on which they can reflect to lead to an important dissipation. This effect is highlighted by Eq. \ref{dissipequa} {which shows that the dissipation} strongly depends on the mass and radius aspect ratio of the radiative core of the star. Since these ratios are equal to zero in fully convective stars, compared to partly convective stars, fully convective stars are then less dissipative.}
\begin{figure*}[ht]
\begin{center}
\includegraphics[angle=-90,width=0.5\textwidth]{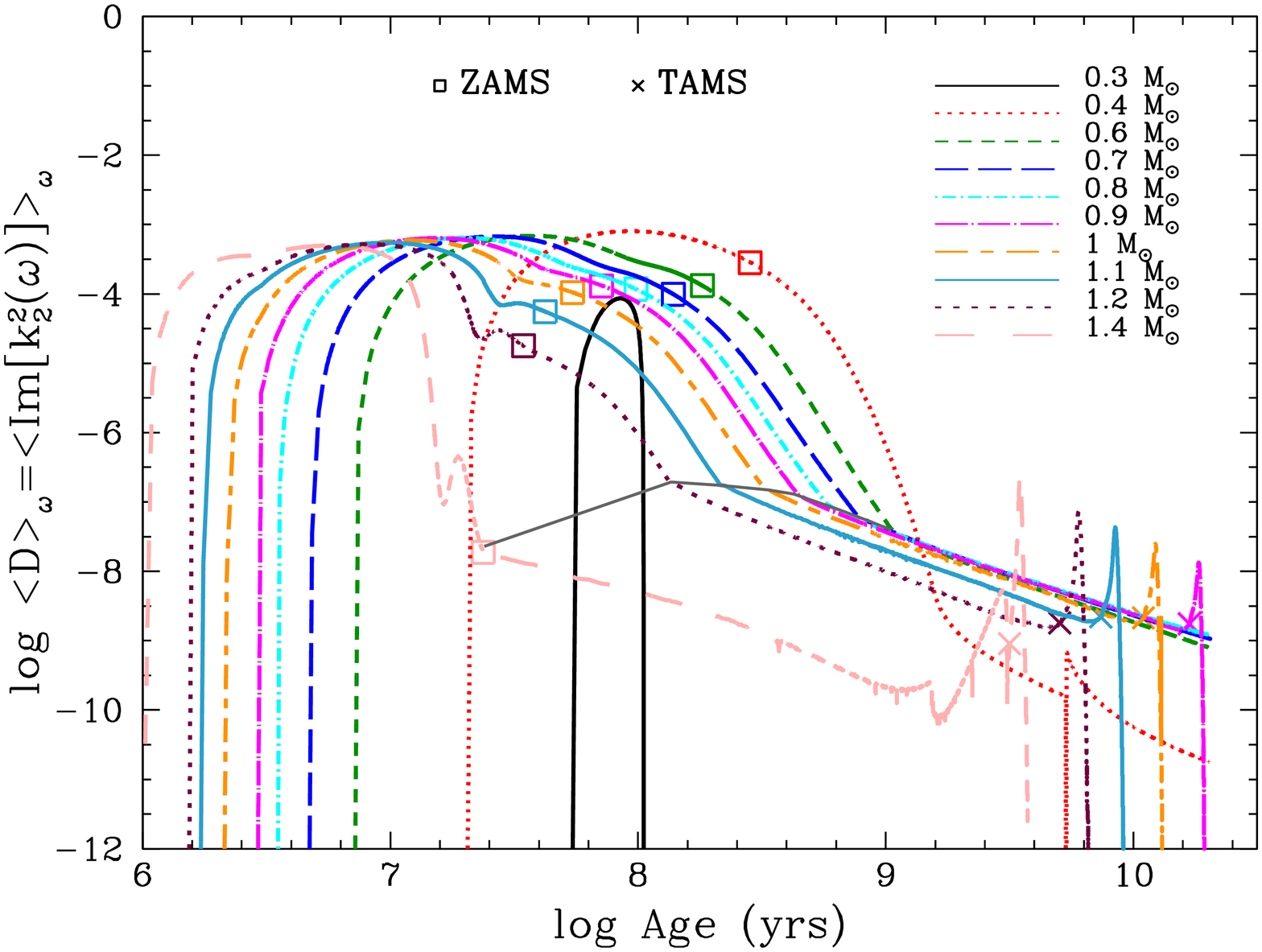} \hspace{-0.5cm} \includegraphics[angle=-90,width=0.5\textwidth]{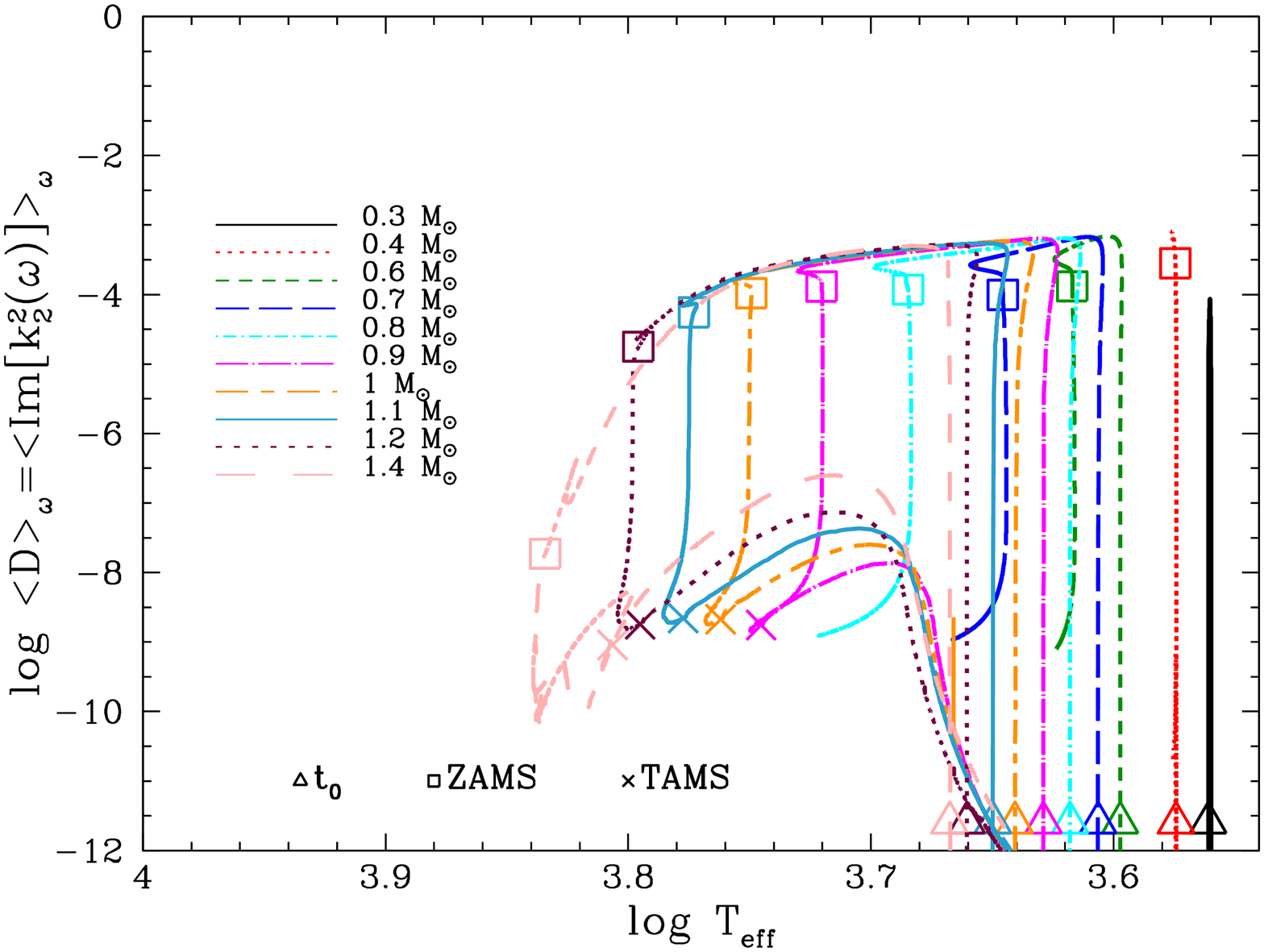} \\
\includegraphics[angle=-90,width=0.5\textwidth]{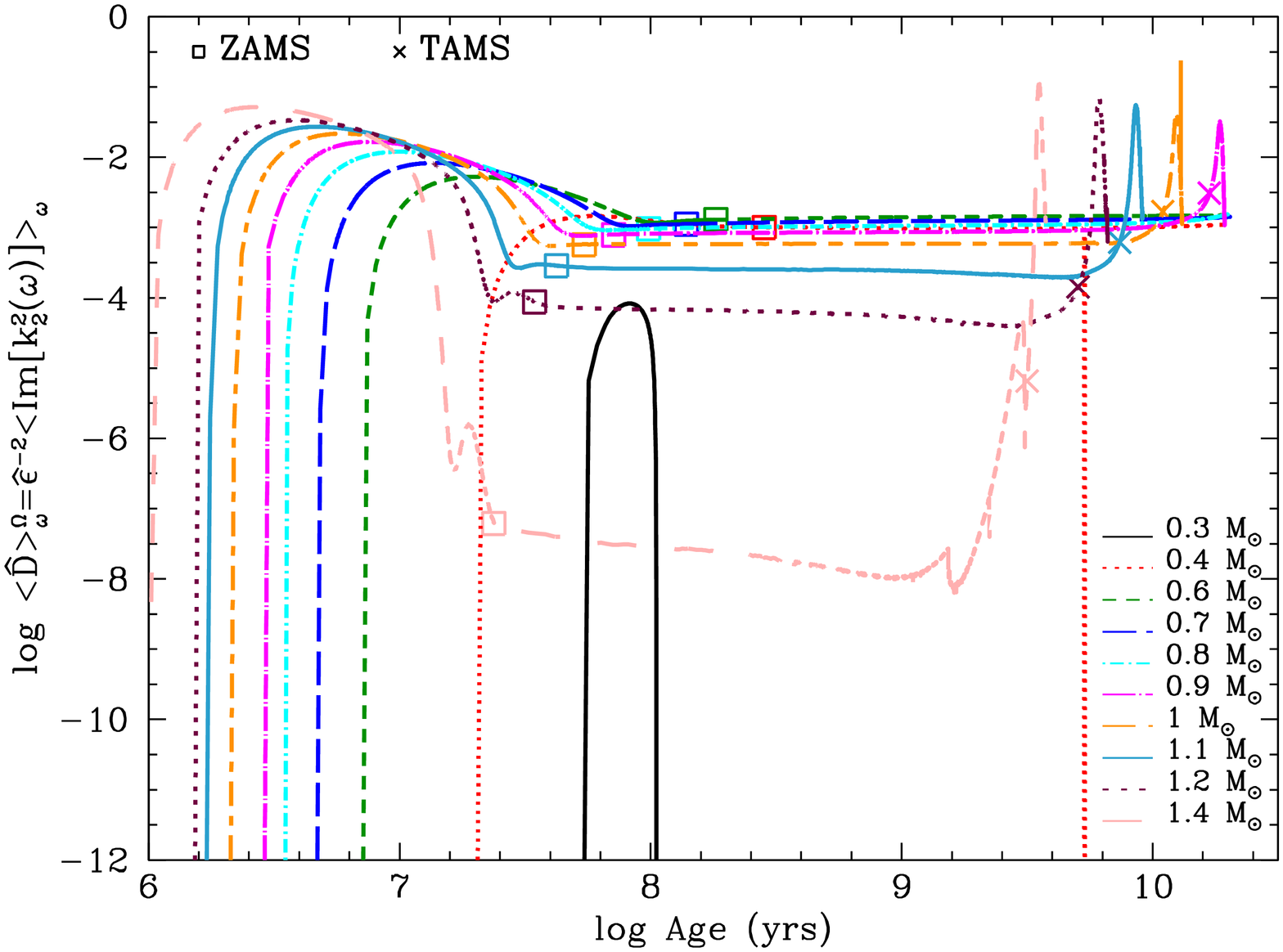} \hspace{-0.5cm} \includegraphics[angle=-90,width=0.5\textwidth]{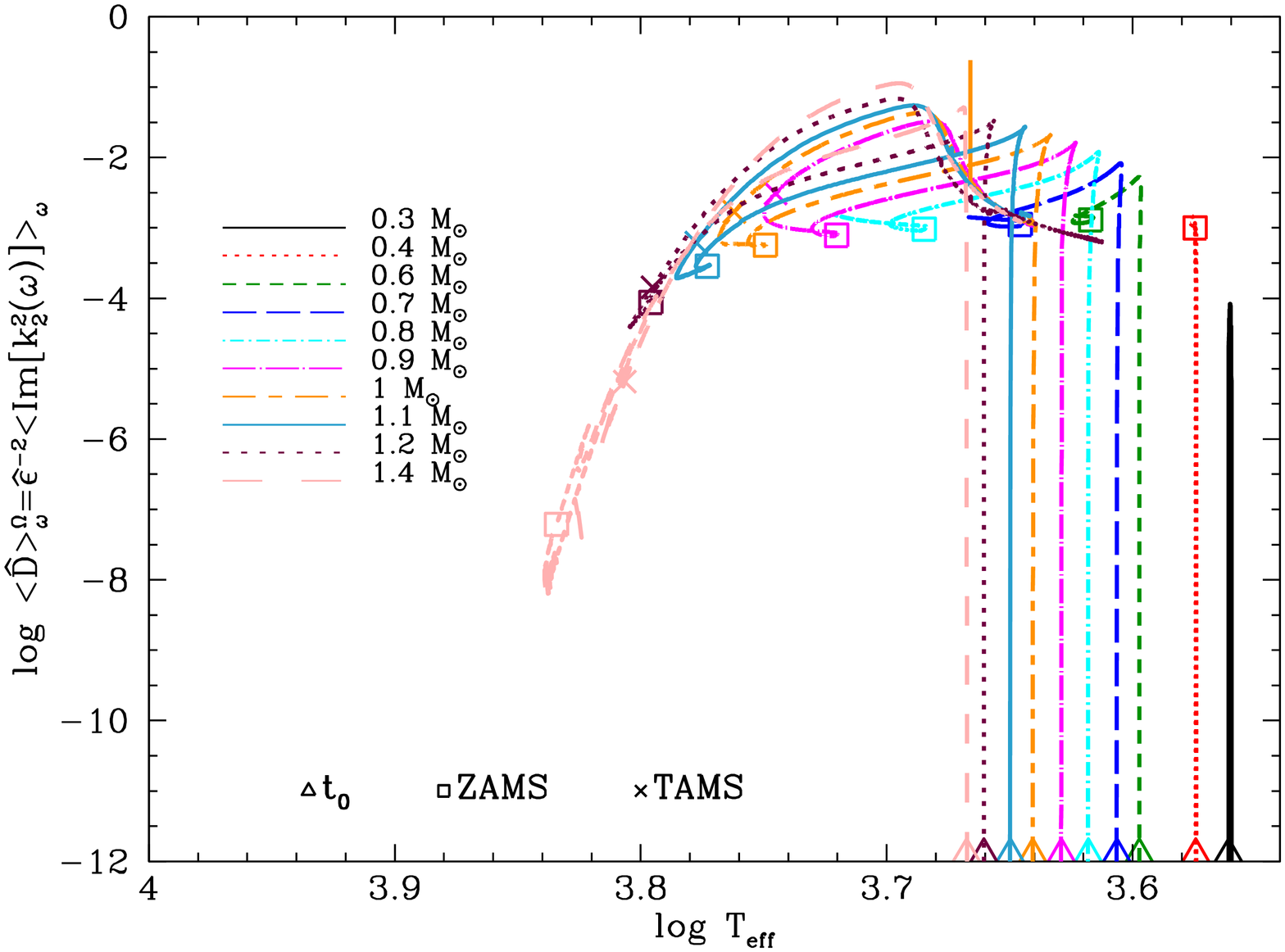}
\end{center}
\caption{{Upper panels:} Evolution of the frequency-averaged tidal dissipation $<\mathcal{D}>_{\omega}=<\rm{Im}\left[k_2^2(\omega)\right]>_{\omega}$, as a function of time (left) and effective temperature (right) for stellar masses ($M_{\star}$) from 0.3 to 1.4 $M_{\odot}$. {Lower panels:} Same as the upper panels but at a fixed normalized angular velocity $<\mathcal{\hat{D}}>_{\omega}^{\Omega}=\hat{\epsilon}^{-2}<\rm{Im}\left[k_2^2(\omega)\right]>_{\omega}$.}
\label{dissip}
\end{figure*}

\subsubsection*{ZAMS-TAMS}

{Even though the stars stay {about 90 \%} of their life on the main sequence (MS), their path in Fig. \ref{chemin} evolves very little during that phase, and remains in the low tidal dissipation region (upper black era) with an almost constant frequency-averaged tidal dissipation. This is due to the fact that the stellar radius and internal structure (thus $\alpha$ and $\beta$) evolve only modestly along the MS. Close to the terminal-age main sequence (TAMS; crosses in Fig. \ref{stellarstructure}) the stellar radius starts to increase more rapidly without noticeable change of the mass of the convective envelope, which induces a decrease of $\alpha$ at almost constant $\beta$.}

\subsubsection*{Evolved phase}

{As the stars evolve along the SG towards low effective temperatures, the stellar radius increases and the convective envelope deepens in both mass and radius. This explains the sharp and rapid decrease of both {the} mass and radius aspect ratio{s} $\alpha$ and $\beta$ (Fig.~\ref{stellarstructure} and \ref{chemin}). At the end of the first dredge-up on the red giant branch (RGB), the convective envelope recedes again in mass, and $\beta$ slightly increases again. Consequently,} during the SG and RGB phase the {evolution tracks in the ($\alpha$, $\beta$) plane} pass {again} through regions of higher dissipation (up to an intensity of log $<\mathcal{D}>_{\omega}^{\Omega} = -2.5$) before heading in the region of very low dissipation towards the RGB {when the stellar radiative core is very small in both mass and radius.} 

\subsubsection*{Summary - Hysteresis-like cycle along the evolution}

With Fig. \ref{chemin} we clearly see the evolutionary path followed by low-mass stars in the ($\alpha$, $\beta$) plane as well as the resulting {evolution of the} tidal dissipation intensity (at fixed rotation). {All {the solar-metallicity models with masses between 0.3 and 1.4~M$_{\odot}$} develop a radiative core {and} follow an hysteresis-like cycle: {the tidal dissipation intensity is first very low, then it increases and reaches a maximum value during the PMS phase before decreasing again at the arrival on the ZAMS; it stays almost constant along the MS, 
and increases again during the first dredge-up phase on the SG and at the base of RGB before decreasing in the upper part of the RGB.}
{As expected, we confirm the results of \citet{Mathis15} and extend the predictions towards more advanced evolution phases.}

\begin{figure*}[ht]
\begin{center}
\includegraphics[angle=-90,width=0.49\textwidth]{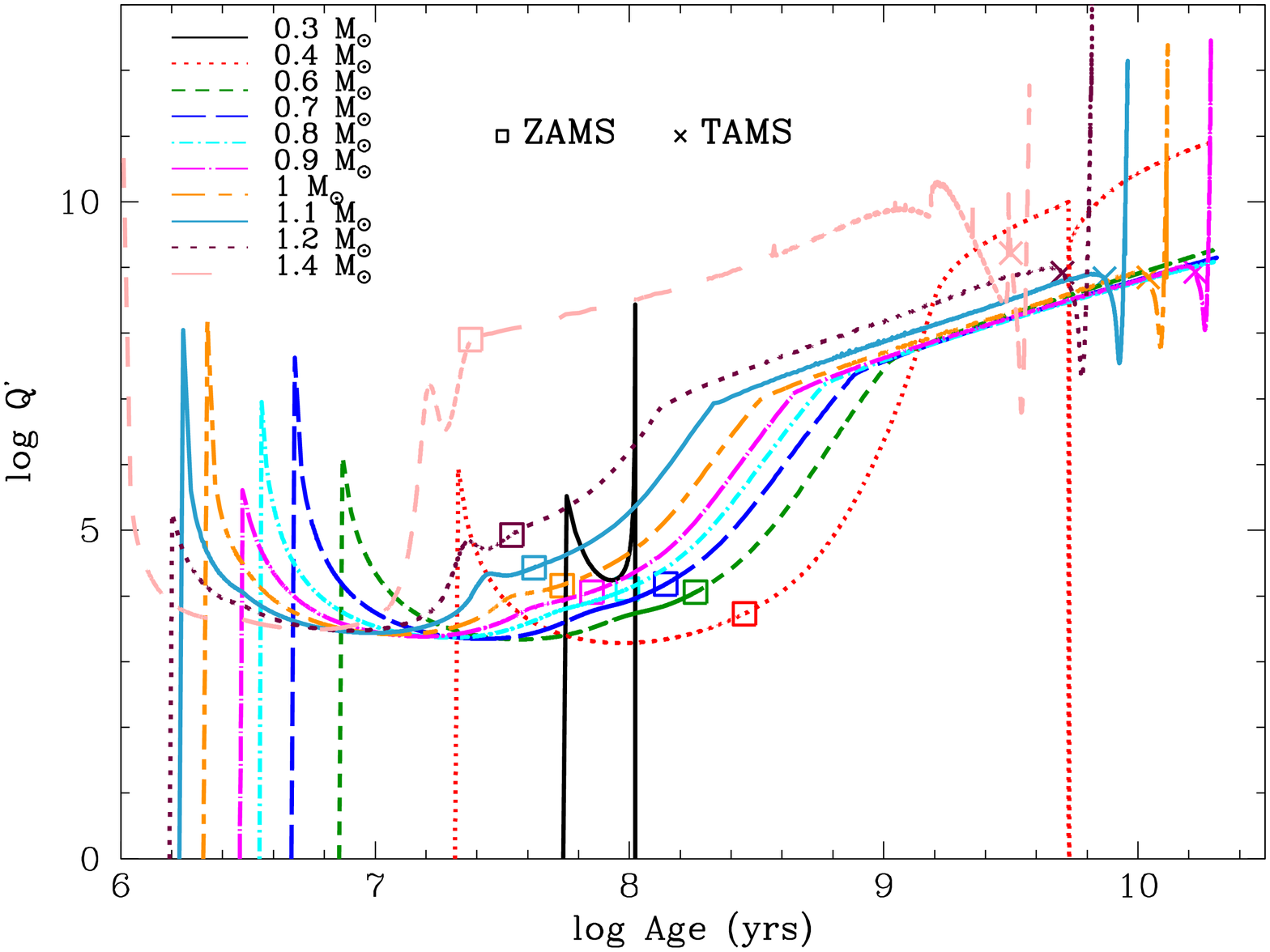} \hspace{-0.1cm} \includegraphics[angle=-90,width=0.49\textwidth]{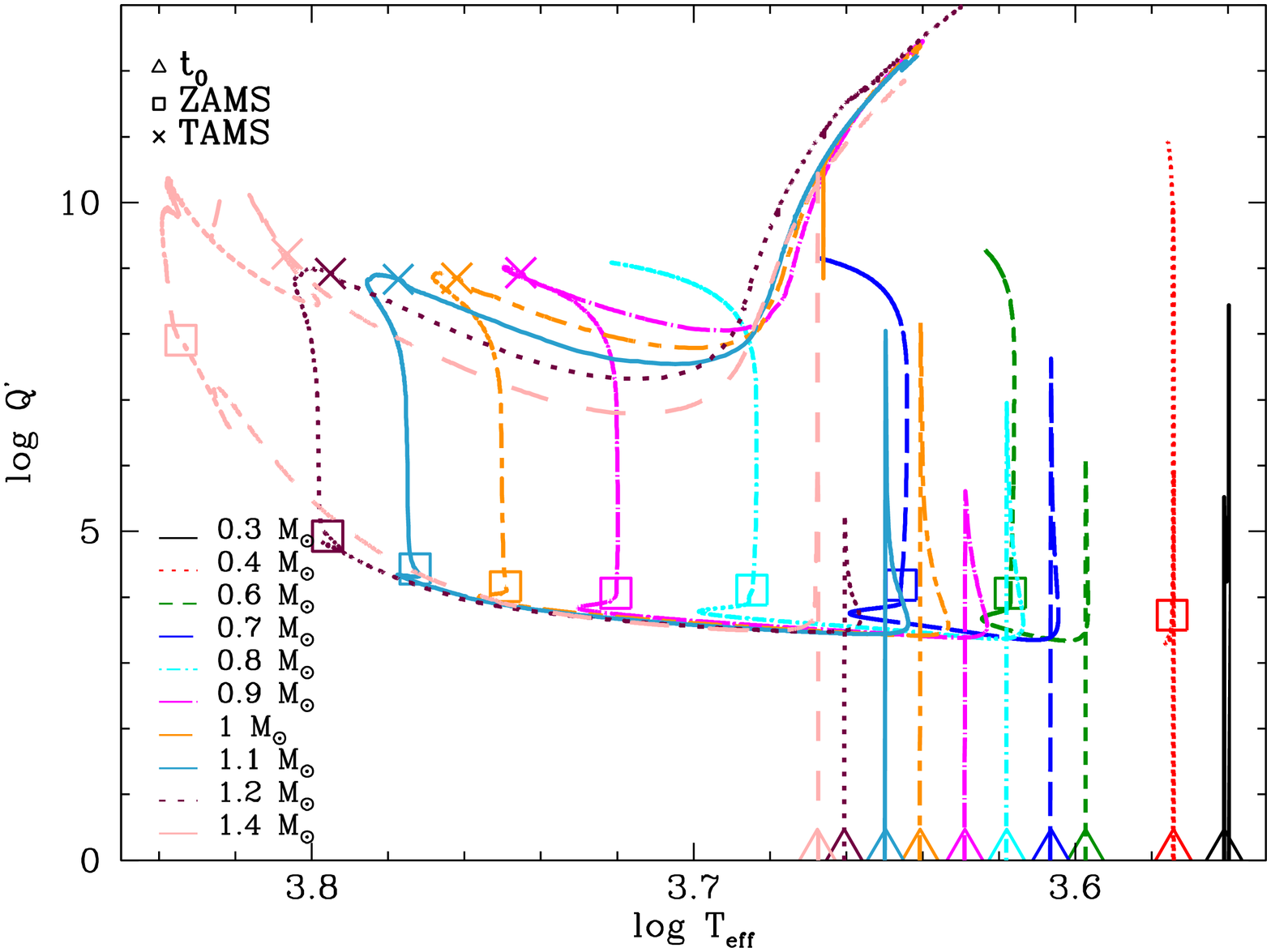} \\ 
\includegraphics[angle=-90,width=0.49\textwidth]{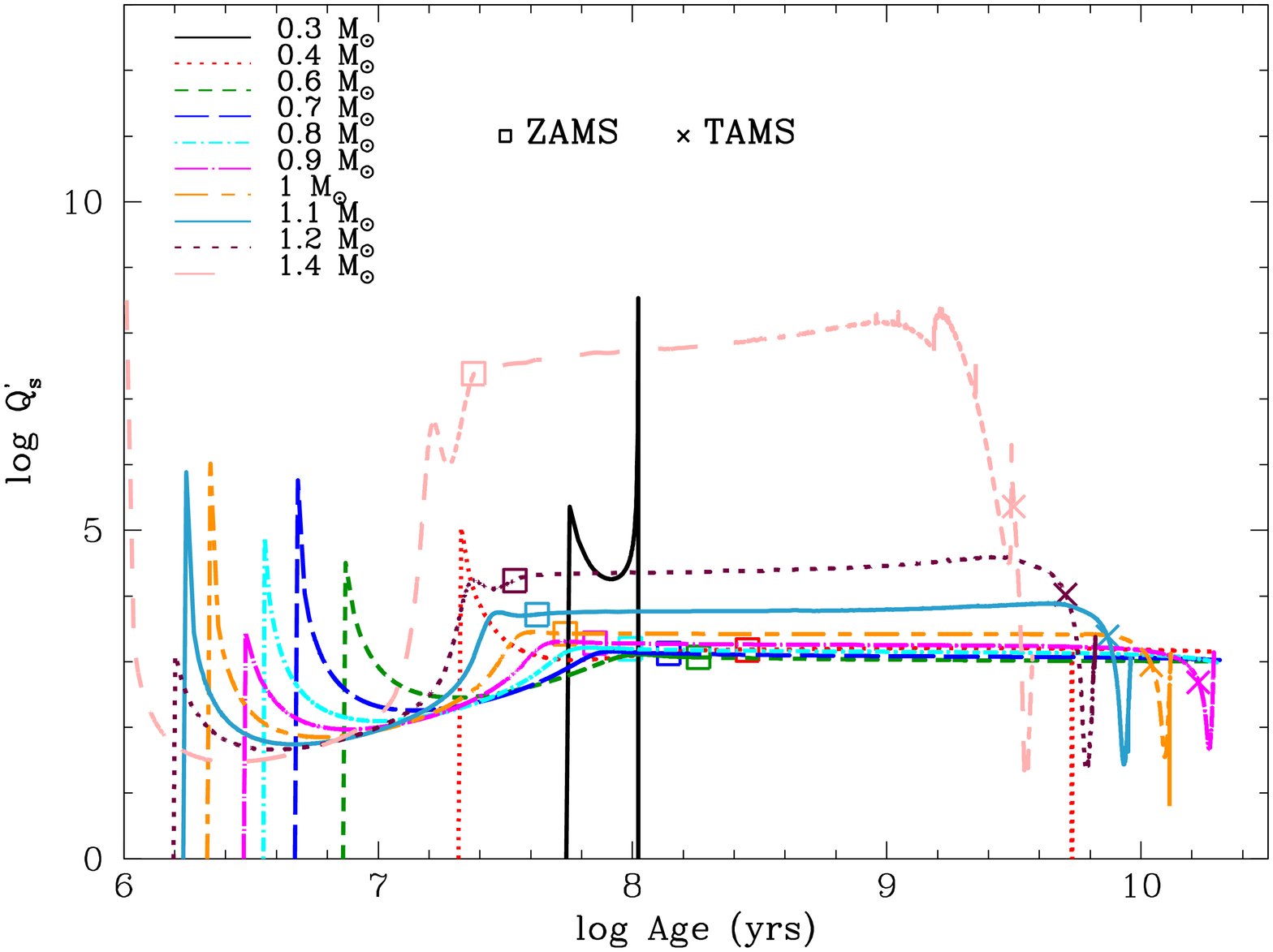} \hspace{-0.1cm} \includegraphics[angle=-90,width=0.49\textwidth]{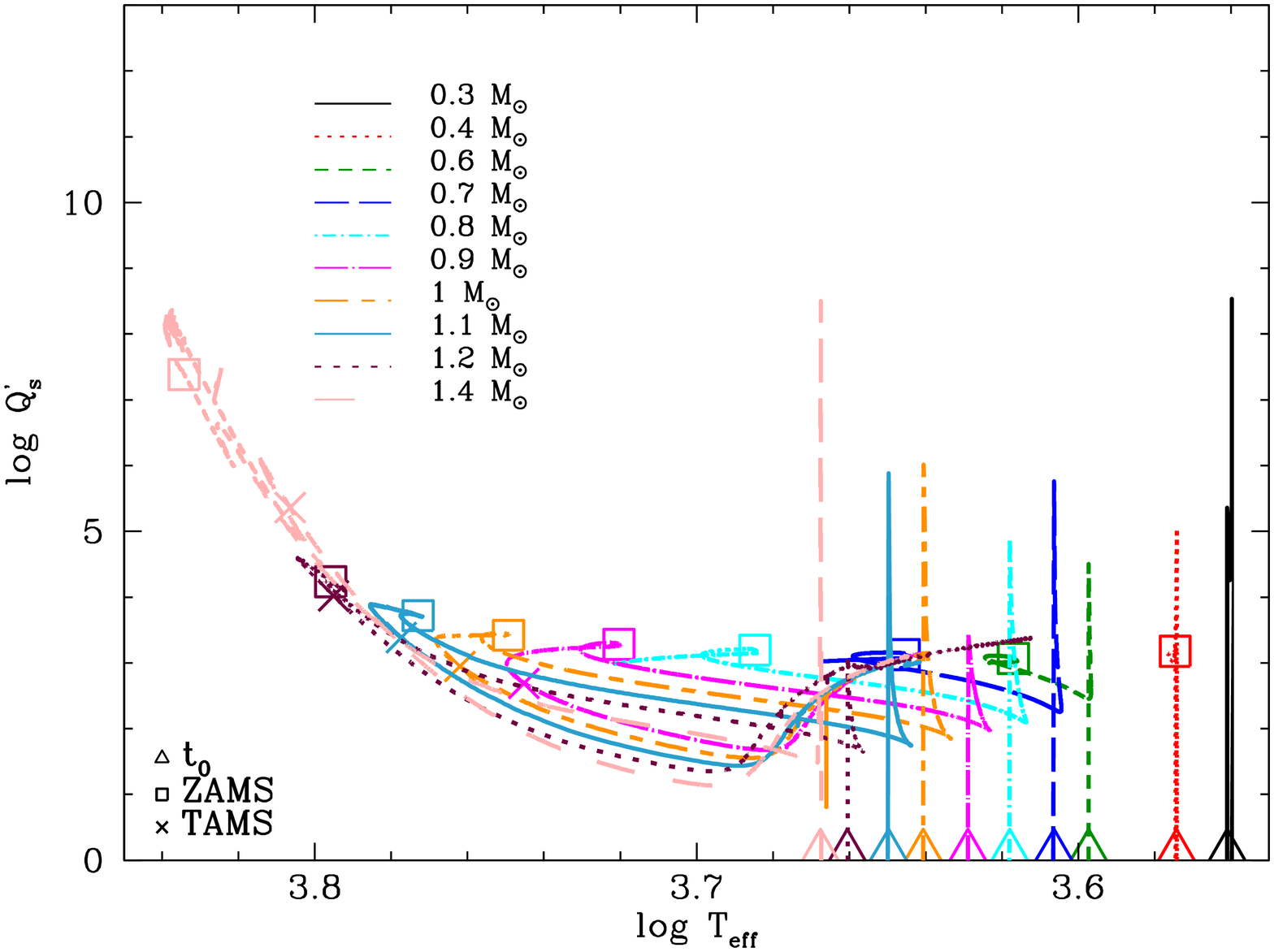}
\end{center}
\caption{{Upper panels:} Equivalent modified tidal quality factor $\overline{Q'}=3/(2<\mathcal{\hat{D}}>_{\omega})$ as a function of time ({left}) and effective temperature ({right}). {Lower panels:} Equivalent  structural modified quality factor $Q_s^{'}=3/(2<\mathcal{\hat{D}}>_{\omega}^{\Omega})$ as a function of time ({left}) and effective temperature ({right}).}
\label{qprime}
\end{figure*}

\subsection{Rotational effect: dissipation and equivalent modified quality factor as a function of time and effective temperature}
\label{rotationaleffects}

{We explore now the effects of the variations of the stellar rotation rate along the evolution on the dissipation and equivalent modified quality factor. As described in \S~\ref{model}, the surface velocity of our stellar models evolves under the action of the secular variations (expansion and contraction), of magnetic braking, and of internal transport processes. We refer to \citet{Amard15} for details. What matters for the present study are the general trends:
After few Myrs on the PMS star is disconnected from its disks, it spins up as its stellar radius decreases up to the ZAMS.
On the MS, the surface rotation decreases continuously  because of the wind braking. 
After the TAMS, the expansion of the stellar radius leads to the continuous decrease of the stellar angular velocity until the star reaches the tip of the RGB.}

{Figures \ref{dissip} and \ref{qprime} show the evolution of the frequency-averaged tidal dissipation and its corresponding equivalent quality factor, respectively. In each of these figures, the upper panels display the full dissipation/equivalent quality factor including structural and rotational evolution effects; for comparison, the lower panels display its normalized version (here normalized to the sun's critical ratio $\hat{\epsilon}$, {see Eq.~\ref{norm_sun}} and \ref{qnormequa}) where rotational effects are filtered out.} 
The left panels of Fig.~\ref{dissip} and \ref{qprime} display the tidal dissipation and equivalent quality factor as a function of time while the right panels show these quantities as a function of the effective temperature {(that is a more physical quantity compared to age).} 
Note that the lower panels of our Fig. \ref{dissip} allow us to recover the results obtained in Fig. 4 of \citet{Mathis15} on the PMS and the MS.

The evolution of the normalized tidal dissipation follows the evolution of the internal structure of the star. In the lower left panel of Fig. \ref{dissip} we retrieve the different regimes followed by the frequency averaged tidal dissipation intensity at fixed rotation {as described in \S~\ref{structuraleffects}}. 
This structural modulation is also found, but inverted, in the evolution of the equivalent structural quality factor as a function of time (see lower left panel of Fig.~\ref{qprime}). {The main effects of relaxing the normalization on the rotation rate are  (i) that the tidal dissipation is lower, on the PMS, by about two orders of magnitude toward lower intensity as $\epsilon \ll 1$, and (ii) that the behavior of the dissipation, on the MS phase, is drastically changed (see Fig. \ref{dissip} and \ref{qprime}) because of stellar spin-down driven by magnetized winds.}

\begin{figure*}
\begin{center}
\includegraphics[angle=-90,width=0.50\textwidth]{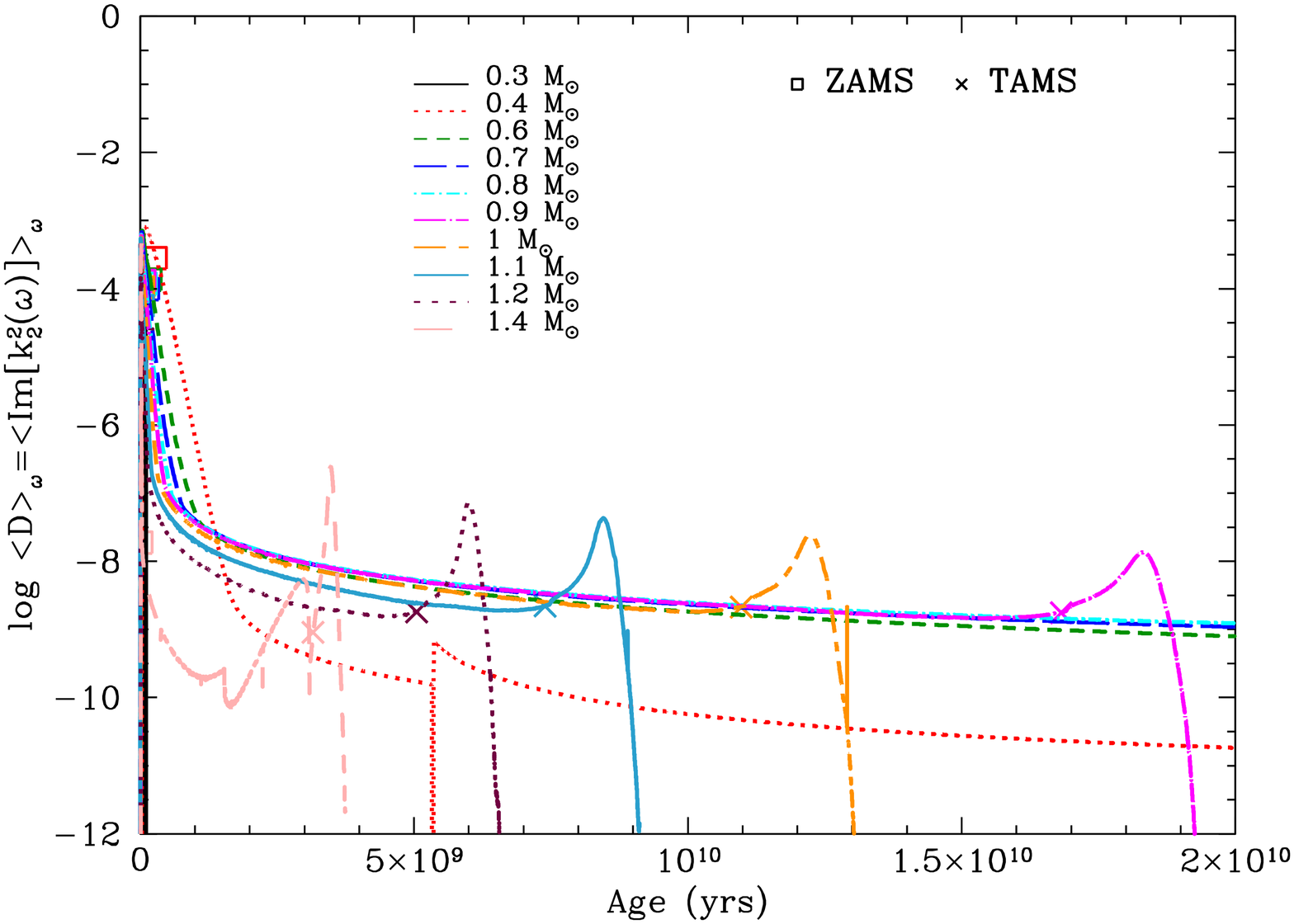}  \hspace{-0.5cm} \includegraphics[angle=-90,width=0.50\textwidth]{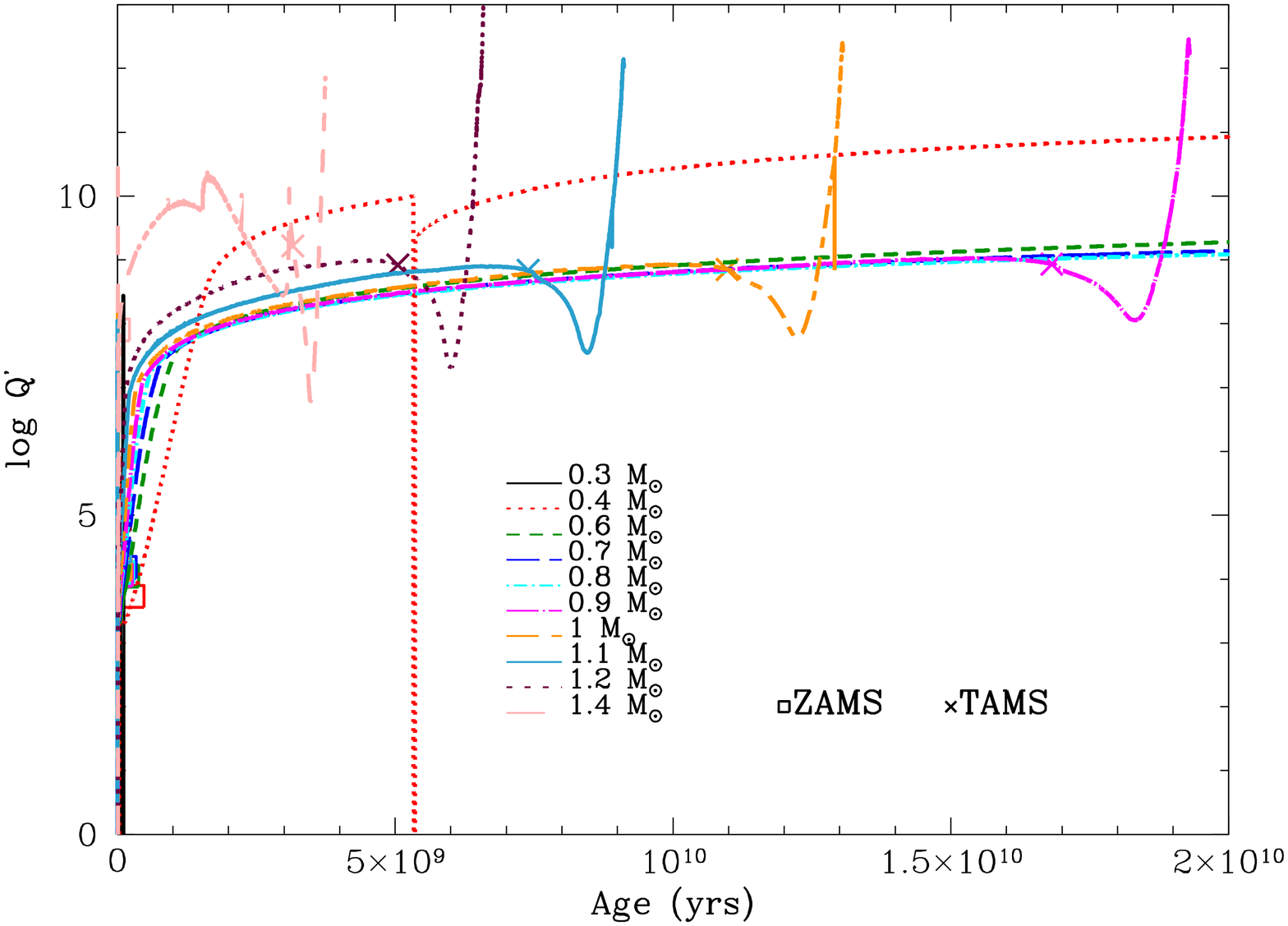}
\end{center}
\caption{Frequency-averaged tidal dissipation $<\mathcal{\hat{D}}>_{\omega}$  ({left}) and equivalent modified tidal quality factor $\overline{Q'}=3/(2<\mathcal{\hat{D}}>_{\omega})$ ({right}) as a function of time in linear scale. The symbol represent: the ZAMS (square) and the TAMS (cross).}
\label{qprimenolog}
\end{figure*}

Even if the rotation rate is evolving during the PMS phase (see upper right panel of Fig.~\ref{stellarstructure}) it has no impact on the behavior of the tidal dissipation (or equivalent quality factor) since the rotation itself is entirely controlled by the stellar contraction (i.e. by the internal structure). {During the Hayashi phase, as the star contracts and its core develops, the tidal dissipation (equivalent quality factor) first increases (decreases) at almost constant effective temperature (see right panels of Figs.~\ref{dissip} and \ref{qprime}). Then on the Henyey phase, the dissipation (equivalent quality factor) reaches a plateau while the effective temperature slightly decreases. {Just before the ZAMS, the dissipation (equivalent quality factor) decreases (increases) as the star slightly expands}.} 

During the MS phase, and as pointed out above, both the mass and radius aspect ratios remain more or less constant. At that point, the internal structure stops to control the evolution of the tidal dissipation (equivalent quality factor). From the ZAMS and up to the TAMS, the tidal dissipation (equivalent quality factor) is controlled by the evolution of the {surface} angular velocity and thus by the extraction of angular momentum (see Sect. \ref{model}). 
As a consequence, the tidal dissipation (equivalent quality factor) continuously decreases (increases) towards the TAMS. Note the stall in this evolution that is due to the transition 
between saturated and unsaturated wind regime \citep[see][and references therein and the gray line in the upper left panel of Fig. \ref{dissip}]{Matt15}. Indeed, this stall in almost all 
rotational tracks is due to a change in saturation regime induced by the saturation of the magnetic field that observationally appears around Ro = 0.1 \citep{Saar96,Saar01,RM12}. 
The effect of this magnetic saturation is to reduce the efficiency of the braking law \citep[see][]{Kawaler88}. During that phase, while the temperature decreases, the tidal dissipation 
(equivalent quality factor) linearly in logarithmic scales decreases (increases) at quasi constant effective temperature. 

Finally, after the TAMS and along the SG and RGB phases, the dissipation (equivalent quality factor) first starts to increase (decrease) as both mass and radius aspect ratio move closer to the island of maximum intensity (see Fig.~\ref{chemin}), and then decreases (increases) as both aspect ratios are strongly reduced by the stellar expansion, leading to structure closer to weakly dissipative fully convective stars. 

{The tidal dissipation and equivalent quality factor are thus strongly affected by the {variations of the rotation rate along the evolution of the star}.} $\overline{Q'}$ reaches maximum values up to $10^{12}$ (for the more massive stars of our sample) from $10^{6}$ to $10^{10}$ yrs while $\overline{Q'}_{s}$ reaches maximum values up to $10^{8}$ during the {same period}. {Compared to the case of fixed angular velocity (\S~\ref{structuraleffects}  and \citealt{Mathis15}}) the variations of rotation along the evolution} lowers the values of the equivalent quality factor and tidal dissipation by four orders of magnitude. This is especially true on the MS when the structure ($\alpha$, $\beta$, $R_\star$) is almost fixed but the rotation rate evolves {(decreases)} significantly. 

The right panels of Fig. \ref{dissip} and \ref{qprime} additionally show the hysteresis cycle followed by the higher mass stars considered here. From the earliest steps of the PMS phase up to the RGB, the tidal dissipation almost achieves a loop by nearly reaching its initial starting point. Note that this hysteresis cycle that is clearly visible in the case of the normalized tidal dissipation is not as pronounced in the case of the non-normalized one because of the action of rotation.

\section{Orbital evolution during evolved stellar phase}\label{orbit_evol}

We showed in \S\ref{results} that the tidal dissipation $<\mathcal{D}>_{\omega}$ (or the equivalent tidal quality factor $\overline{Q'}$) strongly varies from the PMS to the RGB along 
with the structural parameters and rotation rate of the star\footnote{{Note that our tidal orbital evolution models are not now strictly coupled to the stellar evolution models. 
We use grids for the structural tidal dissipation $\overline{Q'}_s$ which come from grids of $\overline{Q'}$ calculated for an initial rotation period of 1.4~day. 
We then compute $\epsilon$ to recover a consistent $\overline{Q'}$ from the evolution of the stellar rotation given by the Equations of \citet{Bolmont12} and \citet{Bolmont16}.}}. 
These variations of several orders of magnitude should have an impact on the tidal evolution of close-in planets.
We investigate here the effect of the tidal dissipation evolution during the evolved stellar phases on the orbital evolution of a 1~$M_{\rm{jup}}$ mass planet. The other phases of tidal evolution have been intensively investigated in \citet{Bolmont16}. In particular, the high dissipation occurring during the PMS phase is responsible for important planetary migration. {Indeed, by including the frequency-averaged dynamical tide formalism of \citet{Mathis15} and \citet{Ogilvie13} in an orbital evolution code, \citet{Bolmont16} pointed out strong outward migration of close-in planets outside the stellar co-rotation radius and inward migration (with the planet that eventually fall into the star) for close-in planet initially inside of the co-rotation radius. With this work, they completely change the conclusion of \citet{Bolmont12} that only used the equilibrium tide formalism.}

The evolved phases correspond here to the SG phase and RGB phase that are represented in Fig.~\ref{hrd}.
The evolutionary tracks can be divided into two parts: the SG phase occurring just after the TAMS (cross symbol in Fig.~\ref{hrd}), which is characterized by a quasi constant luminosity and a decrease of effective temperature, and the RGB phase, which is characterized by an increase of luminosity and a decrease of effective temperature. Figure \ref{qprimenolog} shows the evolution of the tidal dissipation ({left}) and equivalent quality factor ({right}) as a function of time in linear scale for these two late phases. 

{At the end of the MS phase, there is no more hydrogen in the core and the helium core is then deprived of nuclear sources and contracts. The stellar core becomes isothermal (the temperature is insufficient to burn helium) and contracts. Hydrogen burning then migrates into a shell around the helium core. 
The combination of core contraction and shell hydrogen burning leads to an expansion of the stellar radius and to an inflation of the envelope. The SG phase corresponds more precisely to a decrease of the effective temperature at almost constant luminosity, which are a direct consequence of the convective envelope expansion and the core contraction. The tidal dissipation is therefore increased during a short phase as the mass and radius aspect ratio of the radiative core decreases toward lower values and the star crosses again the $(\alpha,\beta)$ plane from top right to bottom left. However, once the maximum is reached, the dissipation sharply decreases close to zero. It results in a ''bump'' seen in Fig. \ref{dissip} and \ref{qprime} between 3 Gyr (1.4 $M_{\odot}$) and 20 Gyr (0.9 $M_{\odot}$). The underlying idea is to know whether this bump has an impact on the evolution of the semi-major axis of a Jupiter mass planet or not.}

Figure \ref{smaevol1msol} shows the evolution of the semi-major axis of a 1 $M_{\rm{jup}}$ planet orbiting a 1 $M_{\odot}$ star with an initial rotation period of 1 day. {The orbital evolutions were computed using Eqs. \ref{Hansena} and \ref{Ts} \citep[see Section \ref{orb_evol_model} and][for more details]{Bolmont16}.} In contrast to previous models {that did not take into account the contribution of the dynamical tide}, this model allows a more complete picture of the stellar dissipation.
\begin{figure*}[ht!]
     \begin{center}

        \subfigure[$1.0 ~M_{\odot}$]{
            \label{model10}
            \includegraphics[angle=-90,width=0.51\textwidth]{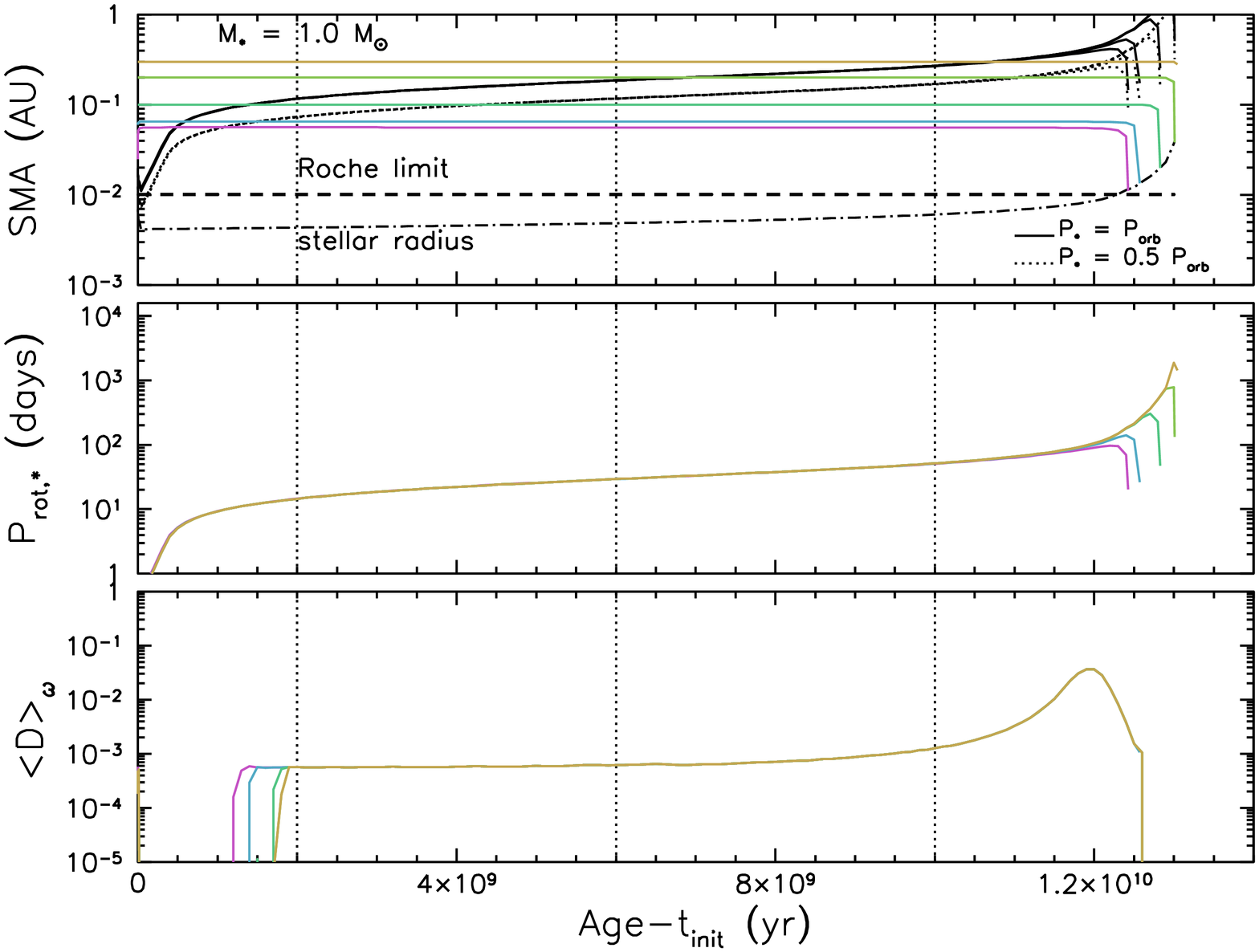}
        } \hspace{-1.4cm}
        \subfigure[$1.2 ~M_{\odot}$]{
           \label{model12}
           \includegraphics[angle=-90,width=0.51\textwidth]{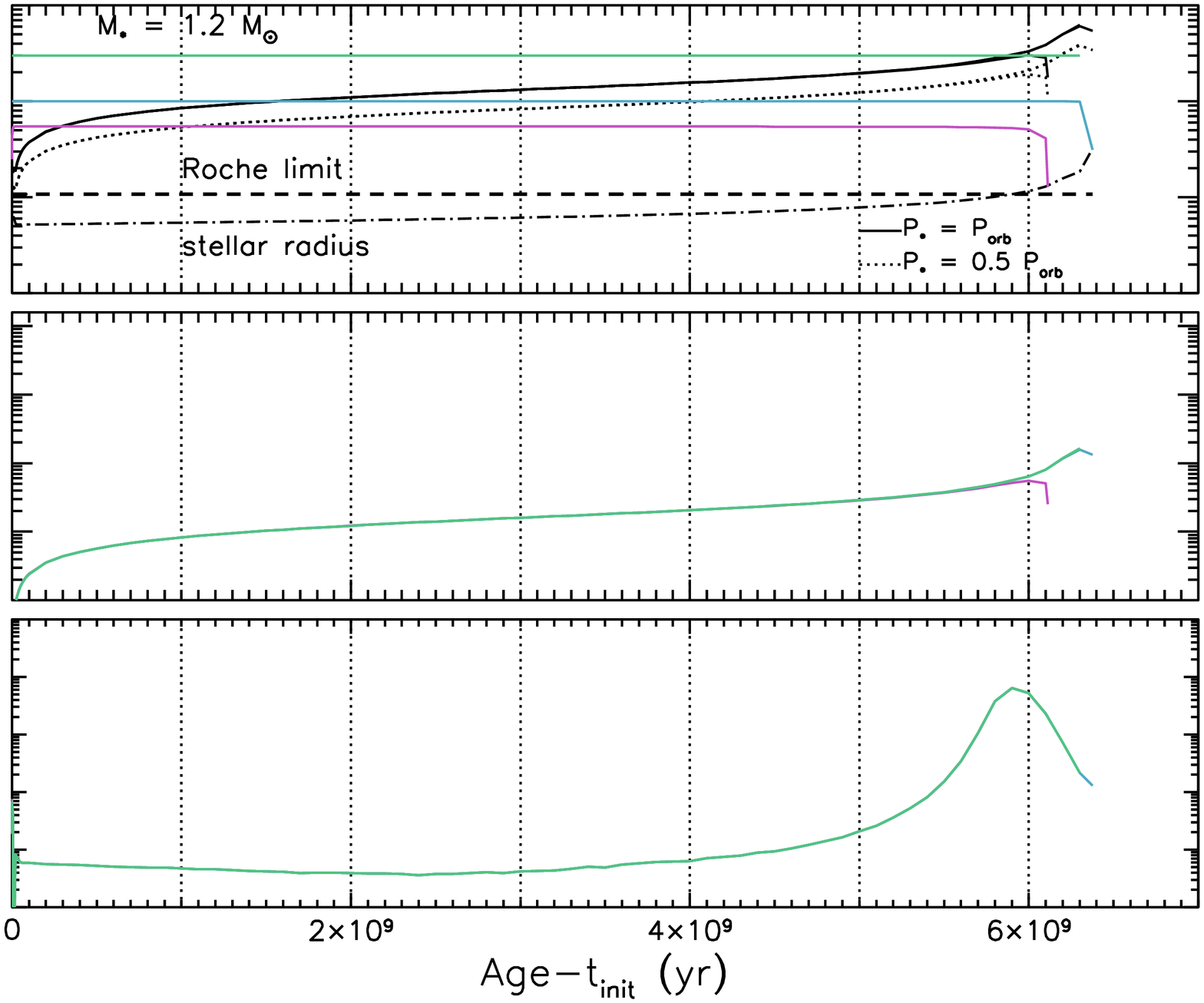} 
        }
    \end{center}
    \caption{Evolution of the orbital distance of a Jupiter mass planet with different initial SMA (top panel), the stellar rotation period (middle panel) and the stellar dissipation factor (bottom panel) during the evolved phases (sub-giant and early RGB) of a 1.0 $M_{\odot}$ ({left)} and 1.2 $M_{\odot}$ ({right}) star . {Top panel}: the orbital distance of the planet is represented in full colored lines. The full black lines correspond to the co-rotation distance ($P_{\rm orb} = P_{\star}$), and the dotted black lines to $P_{\rm{orb}} = 0.5~P_{\star}$ delimiting the region where the dynamical tide operates. The long dashed black line corresponds to the Roche limit and the dashed-dotted line to the stellar radius. {Middle panel}: the surface rotation period (in days) of the host star.  {Bottom panel}: the stellar dissipation  $<\mathcal{D}>_{\omega}$. The time on the x-axis is given from an initial time $t_{\rm init}$, which corresponds to the time of the protoplanetary disk dispersal. This initial time is taken to be 5~Myr.}
   \label{smaevol1msol}
\end{figure*}
{We again recall the reader that the rotational evolution of the stars is estimated using the braking law from \citet{Bouvier1997}. Even if the parametrization of the \citet{Bouvier1997} braking law is theoretically only valid for stars between 0.5 and 1.1 $M_{\odot}$, in our model the rotational evolution of the 1.2 $M_{\odot}$ star is consistent with the expected behaviour of a 1.0 and 1.1 $M_{\odot}$ \citep[see][]{GB15,Amard15} with a surface rotation rate that reaches a 10 days period between 1 and 2 Gyr.}

Figure \ref{smaevol1msol} ({left} and {right}) shows that this bump has no effect on the evolution of the semi-major axis (hereafter SMA) of the planets.
There are two main reasons for this behavior.
The first reason is that planets susceptible to experience this bump in dissipation are located {too} far away for tides to impact them significantly. 
Indeed, they have to be in the region in which they excite the inertial waves in the convective envelope (i.e., where $P_{\rm orb}> 1/2 P_\star$, $P_{\rm orb}$ is the orbital period of the planet and $P_\star$ is the rotation period of the star) at the moment of the bump.
The very close-in planets cross the limit $P_{\rm orb}= 1/2 P_\star$ too early in the evolution of the system. 
Only planets farther away than $\sim 0.3$ AU are still in the dynamical tide region at the time of the bump.
The second reason is that the star has slowed down so significantly that at the age of the bump (around 12~Gyr for $1~M_\odot$ and $6$~Gyr for $1.2~M_\odot$), its rotation is very slow ({period} of the order of 100~days).
\citet{Mathis15}, \citet{Bolmont16} and Fig. \ref{qprimenolog} show that the dissipation in the star decreases as the star spins down, so that at the age of the bump the dissipation is actually very low and does not impact the orbital evolution of the planet.
Figure \ref{smaevol1msol} ({left}) actually shows that due to the stellar spin down, the bump is actually not visible.
However, Fig. \ref{smaevol1msol} ({right}) shows that for $1.2~M_\odot$ the bump is visible.

Figure \ref{smaevol1msol} ({left} and {right}) also show that when falling onto the expanding star the planets make it accelerate significantly.
For instance, we find that a planet at 0.05 AU at 4.5~Gyr, induces a decrease of the rotation period from $\sim$180~days (the value it would have without planets) to 20~days.
This corresponds to an increase in surface velocity from 0.68~km.s$^{-1}$ to 6~km.s$^{-1}$.
\citet{Privitera2016} also studied the influence of planet engulfment on stellar spin up for stars and find the same type of behavior. 
However, their study was focused on higher mass stars ($>1.5~M_\odot$) and even later stages than in this work.
They also took into account the mass loss from the star and the head wind planets feel due to the ejected matter. 
{We did not take into account these phenomena because the mass loss for the stellar mass range and the phase we consider is not as important as for the objects they study.}

\medskip

We consider here Jupiter-like planets, however this model could be used to investigate the future of the Earth as the Sun expands \citep[e.g., as was done in][]{Schroder2008}.
Nevertheless, the computed evolutionary models do not go to sufficiently advanced phases for such a small-mass planet at 1~AU to be impacted, and we do not here take into account the effect of the mass loss from the star on the orbital evolution of the planets, which begins to be important at more advanced stages and which it has a non-negligible effect on the orbital distances of planets. 
Here, our $1~M_\odot$ model stops on the RGB, at an age of 13~Gyr when the radius of the star is of about $0.04$~AU. 
Figure \ref{smaevol1msol} ({left}) shows that even a Jupiter-mass planet at $0.3$~AU is only just starting to be influenced by the tides at that age.
This has several consequences: 1) when an Earth-mass planet is susceptible to be influenced by the stellar tide (when the radius is large enough, see Eq.~\ref{Ts}), the star has spun down enough so that the planet would not evolve because of the dynamical tide but the equilibrium tide, 2) even if the planet evolved due to the dynamical tide, the structure of the star would be such that the dynamic tide would be very weak (due to the huge size of the convective envelope).
We would therefore not expect our model to change what has been done on the future of the Earth as the Sun becomes a red giant.  

\section{Conclusion}
\label{conc}

{We extend the analysis of \citet{Mathis15} of the evolution of the dissipation of tidal inertial waves propagating in the convective envelope of low-mass stars from the PMS to the RGB tip. We take into account for the first time the variations of both stellar rotation and internal structure. As a first step, we assumed for the tidal dissipation model a simplified bi-layer stellar structure where the radiative core and the convective envelope have averaged densities.  This allows us to  obtain an analytical expression for the frequency-averaged tidal dissipation, which provides us a reasonable order of magnitude of the dissipation in stellar convective envelopes as a function of their depth, mass and rotation.  In  forthcoming works, the strong frequency-dependence of the dissipation of tidal inertial waves \citep{Ogilvie07} and the radial variation of the density,  which could both affect the strength of tidal friction should be taken into account.  However, numerical modelling will become more complex and heavy. Our approach thus constitutes a first and necessary step to explore a broad parameter space for planetary systems and their host stars. We use these new predictions for tidal dissipation to generalize the work of \citet{Bolmont16}; we follow in particular the orbital evolution of close in planets during advanced stages of stellar evolution.}

By coupling the stellar evolution code STAREVOL to the frequency-averaged tidal dissipation and equivalent modified tidal quality factor prescription of \citet{Mathis15} and \citet{Ogilvie13}, we have shown in this work that stellar evolution is crucial in tidal interaction modeling. {Indeed, both the stellar structural evolution (through the {{presence of a radiative core}, which can enhance the tidal excitation of inertial waves} \citep[e.g.][]{Ogilvie07}), and rotational evolution ({through the scaling in $\epsilon^2$ of the dissipation}) strongly affect the evolution of the tidal dissipation and the corresponding equivalent tidal quality factor.} 

While the evolution of the stellar structure controls the evolution of the dissipation during the PMS phase, rotation governs its evolution during the MS phase. Indeed, during the PMS phase the rotation itself is {restrained} by the evolution of the internal structure through the stellar contraction while during the MS phase, the internal structure remains more or less constant leaving room for a rotational modulation of the dissipation via the wind braking mechanism.  

Thanks to the coupling between STAREVOL and the frequency-averaged tidal dissipation prescription, we can provide the community with an online tool\footnote{https://obswww.unige.ch/Recherche/evol/starevol/Galletetal17.php} in which we will provide tidal dissipation and equivalent quality factor's evolution for each stellar masses considered in this work. 

This grid can be used to better constrain {models dedicated to the study of} the orbital evolution of planetary systems. {More specifically it could be used either by ongoing and past space missions such as CoRoT and {\it Kepler}/K2 \citep{Corot,Kepler,K2} and future space and ground-based observatories such as CHEOPS \citep{CHEOPS}, TESS \citep{TESS}, PLATO \citep{Plato}, and SPIRou \citep{Spirou}.

{In this paper, and based on the work of \citet{Bolmont16}, we explored the orbital evolution of close-in planet during evolved stellar phases (namely the RGB). Using the new predictions of tidal dissipation {for these evolution phases}, we pointed out that the bump observed during the RGB in the frequency-averaged tidal dissipation {for the dynamical tide} has no effect on the semi-major axis of close-in Jupiter mass planets orbiting a 1 and a 1.2 $M_{\odot}$ stars.

{Thanks to these combined developments we have now access to a complete model to follow both the evolution of the tidal dissipation and of the planetary semi-major axis. Indeed, we possess a simplified but robust theoretical prescription for the dynamical tides, we can follow the internal structure and rotation rate of the star thanks to STAREVOL as well as the planetary orbital evolution using the {secular} code from \citet{Bolmont16}. We can thus use these results so as to perform modeling of planetary architecture and planetary population synthesis.} 

{The next step will be to directly couple stellar evolution to orbital evolution code using the frequency-averaged tidal dissipation formalism to fully study the possible retro-action of tides and magnetic interactions on the internal structure and rotation of both stars and planets (for instance, using the prescriptions derived by \citealt{Strugarek14}, \citealt{Strugarek16}, \citealt{Bolmont16} and \citealt{Gallet17}). {In addition, we will take into account realistic density profiles in the convective envelope of low-mass stars (e.g. from the STAREVOL code). This development will be done by solving the full linearized spectral equations derived by \cite{Ogilvie13} for densities that vary with radius.} Finally, in this work we only included the dissipation of tidal inertial waves inside the convective envelope of rotating low-mass stars. The next step will be to extend this analysis to tidal dissipation inside the radiative core of these stars \citep{Ivanov13,Guillot14}, hence completing the present partial physical description. Other dynamical processes such as the effects of differential rotation on tides should also be taken into account \citep[e.g.][]{Favier14,Guenel16}. {In this framework, frequency-averaged and frequency-dependent dissipation should be considered.}

\begin{acknowledgements}
The authors would like to thank the anonymous referee for his/her constructive comments. F. G., C. C., and L.A. acknowledge financial support from the Swiss National Science Foundation (FNS) and from the SEFRI project C.140049 under COST Action TD 1308 Origins. E. B. acknowledges that this work is part of the F.R.S.-FNRS ``ExtraOrDynHa'' research project. S.M. and E.B. acknowledge funding by the European Research Council through ERC grant SPIRE 647383. This work was also supported by the ANR Blanc TOUPIES SIMI5-6 020 01, the Programme National de Plan\'etologie (CNRS/INSU), the  Programme National de Physique Stellaire PNPS (CNRS/INSU), and PLATO CNES grant at Service d'Astrophysique (CEA-Saclay).

\end{acknowledgements}

\bibliographystyle{aa}
\bibliography{Reference}

\end{document}